\documentclass[12pt]{article}
\usepackage{amsfonts}
\usepackage{amssymb}
\usepackage{amsmath}
\newcommand{\n}{\smallskip} 
\newcommand{\G}{\Gamma}

\renewcommand{\d}{\delta}
\renewcommand{\b}{\beta}
\newcommand{\g}{\gamma}

\newcommand{\grad}{{\rm grad}}
\newcommand{\e}{\epsilon}

\newcommand{\ar}{\longrightarrow}

\newcommand{\la}{\lambda}
\renewcommand{\a}{\alpha}
\begin{document}
\title{Course of lectures on physical constructivism}
\author{Y.I.Ozhigov \thanks{Creating of this course is supported by NIX Computer Company, Grant \# F793/8-05.}\\[7mm]
Moscow State University \\
of M.V.Lomonosov, VMK Faculty,
\\ Institute of Physics and Technology RAS
} 
\maketitle

\begin{abstract}
In this course, I talk about the source of mathematical constructivism and its role in the future development of theoretical physics. I describe what physical constructivism is and why it is necessary for the penetration of exact methods of theoretical physics to the area of complex systems, which formally belong to the others natural disciplines. I describe the concrete heuristic for the creating of models of constructive quantum theory – the method of collective behavior. I represent the constructive viewpoint on the quantum computer, which treats it as the model object of many particle quantum physics, and the practical recommendation concerning its building. Due to the known inertia of the educational system constructive methods in mathematics remains mostly unknown to the wide physical community, and I hope that this course will stimulate the interest to the new possibilities, which these methods open. These possibilities represent interest to natural scientists and for programmers as well. 
\end{abstract}
\newpage

\tableofcontents
\newpage

\section{Lection. Hilbert program and its fate}

The strongest source of optimism in the job in science is the perception of its unity. This unity has the certain embodiment in the form of the priority of the exact methods over the unclear reasoning that we can formalize as the leadership of the mathematical methods over all area of natural knowledge. The gradual but continuous progress of mathematics to the end of nineteenth century led David Hilbert to his famous program, which core was the thesis about the necessity of the axiomatization of natural knowledge, and the representation of it in the form of some deductive theory. He expressed this aim approximately as follows: “At first we put into order mathematics (e.g. prove its consistency, and, correspondingly, its absolute reliability). Then we take up physics and represent it in the form of mathematical theory. Then we do the same with chemistry, and after that with biology.” His program contained the various concrete mathematical problems, which were solved in the different times, but the most valuable in Hilbert program was just its core, which expressed the factual plan of conquest the natural sciences by mathematics.

Hilbert program cracked already at the first stage: it failed to manage with mathematics itself! In the thirties of nineteen century, Kurt Gedel proved that in any factually consistent theory the sentence about its consistency is improvable in it. It means that we cannot in principal guarantee that the simple development of mathematics in one day will not lead us to the destroying of its main parts in the form of suddenly appeared contradiction. The realizing of the real sense of Gedel result took some time but it turned many to the shock. It seemed that mathematics has lost the status of fundament and nothing could return the perception of reliability, which comes with an exact proof of theorem, where each step is clear and causes no doubts. 

For us the most important conclusion from the crack of Hilbert plan is the deep understanding of the unity of mathematics with experiments. An experiment represents the integral part of mathematics itself, and just it serves as the source of certainty in the correctness of results even in the case when these results seem abstract! We can put it more exact: physics and mathematics represent the same discipline and their division is no more than the question of convenience in the organization of science, at least in the pre computer era. This thesis is important for the following and I express it more concrete. There is no what somebody call the physical sense independent of mathematics. Just the mathematical apparatus determines what has and what has no the physical sense. I propose the listeners to think about it themselves, and hope that the following material will give the additional arguments for this. 

We thus see that the realization of the plan of axiomatization of natural sciences even did not reach the border of the real physics. How can we estimate it? Is any plan of joining of the natural sciences the fiction, which does not deserve the attention and efforts at all? The main criterion in science differing it from the other forms of mental activity, was always the possibility to predict the happening of events. I propose you to compare the level of reliability of predictions in the area with the developed mathematical methods, for example, in space mechanics in the scale of Solar system, and the prediction of behavior of a living thing, to which mathematical methods are applicable in much lesser degree. The difference is so huge that the connection between the degree of applicability of mathematics and the reliability of predictions becomes evident. Factually, the all theory in the natural sciences is built around the most efficient applications of mathematical methods.

We have no right to refuse from Hilbert program because it would equal to the capitulation of the science at all. It is also doubtless that the aim is just in the complex system dynamics, e.g. the elucidation of processes, which formally belong to chemistry and biology. The question then concludes in the choice of appropriate means for the realization of this aim. It makes us to pay the special attention to the mathematical apparatus. Can the axiomatic method cope with the functions required from this apparatus? The actuality of this question now follows from that in the modern science deals with complex systems where the limitations caused by Gedel type results play the key role. Factually, the axiomatic methods discredited itself already in the crisis of mathematics we mentioned. We then cannot even hope that it can rehabilitate in the applications to the real physics; it can serve as the auxiliary tool only, connected with the systematization of existing knowledge, but cannot be the working tool of the theory. The perception of unreliability of the axiomatic approach leads to the decreasing of enthusiasm in the fundamental substantiation of the observed phenomena among the wide mass of investigators, and to the appearance of surrogate specialized theories of this or that phenomenon. In any case, the axiomatic method is connected with the abrupt decreasing of the predictive force of the theory in its application to the more and more complex systems, so that when we approach the borders of biology this force becomes practically zero. 

We then see that the possibilities of the traditional mathematics in the applications to natural sciences are close to the exhausting. It brings the idea of the modification of mathematical apparatus, for the work with the complex systems. One must clearly understand that it does not mean the replacement of mathematics by the kind of philosophy, or something of that sort, but the modification of mathematics itself, which logically follows from its crisis after Gedel theorems. 

In the order to understand the new adventures, which this modification brings, we must get acquaintance with the main theses of mathematical constructivism, which we intend to use instead of the standard mathematics. 

\subsection{Constructive mathematical logic}

Fortunately, the understanding of the dramatic situation in mathematics after the crisis led mathematicians to very fruitful turn, which is called constructivism. The principal exit from the crisis found in mathematics will soon spread its influence to physics, and I will speak about it. This solution is the constructive mathematics, and we take up it now. Constructivism limits the sphere of applicability of mathematical abstractions by the exact computational procedures. It treat as right only such logical formulas of the form $A$ or $B$ for which we can point what is right exactly: $A$ or $B$. The sentence $A$ or not $A$ is not the universal truth in constructivism. It is the sense of the constructive mathematical logic, or intuitionism. Founders of intuitionism were Kronecker and Brower. The first of them disputed about intuitionism with Hilbert who could not accept the severe limitations that intuitionism imposes to the logical constructions: there are no proofs of the pure existence in it.  

Hoverer, already in that times the magic connection between the constructive mathematical logic and quantum theory became evident. John von Neumann and Garry Birkgoff put forward the thesis that the intuitionism is more appropriate for the description of quantum theory than the classical mathematical logic. Their arguments rest on that the constructive understanding of logical sentences corresponds to the extraction of the information about quantum states via its measurement. For example, the formula ($A$ or $B$) in its constructive treatment means that we have performed the measurement and know in what state exactly our system is: in $A$ or in $B$. 

There is very important for physics concept of pluralism in constructive mathematical logic that is the plurality of the logical sentences understanding. Models of logical intuitionistic theories, in contrast to their classical analogs, contain no absolute true estimations. These estimations may be not only the “truth” or the “false”, but also uncertain, e.g. “not known”. In the last case, this estimation can change in time. In the models of intuitionistic theories, there is the principal parameter: the time. It puts the constructive logic in the particular relations with physical experiments, in comparison with the classical logic.\footnote{I personally have learnt about the work of Neumann and Birkgoff, written in thirtieths years of twentieth on the seminar of A.G.Dragalin. Just Albert Grigorievich explained me the meaning of the constructive ideas of A.A.Markov-young whose lectures I listened. However, for the understanding of the real value of constructivism I still had to overcome the long way in physics.}

The pioneer paper of Neumann and Birkgoff had no big continuation. Theoretical physics uses not the logic directly, but the computational instrument, corresponding to this logic; in classical case, it is the solution of differential equations, integration, differentiation, group representations, etc. Any logical interpretation without the obtaining of new results has the same secondary value as any interpretation of physical theories at all: it only helps to understand already obtained results but gives no new ones. 

\subsection{Algorithm theory}

The appearance of constructivism in mathematical logic was immediately stimulated by the crack of Hilbert program and the axiomatic method. Nevertheless, in that time the main notion of constructivism – algorithm yet has not been defined. Turing, Post and Markov-young did it independently. Many bodies connect the necessity of the formal definition of algorithms with the need to prove the absence of an algorithm with some certain property. There is no doubt in the value of such distinguishing with the old “not algorithmic” world. However, the exact definition of algorithms has also the positive dimension: it concerns the possibility to compute something practically. Of course, many scientists quickly understood the interest for physics (in our country there were Tihonov and Samarsky, the row of their pupils) and the rapid development of what is now known as computational methods. These methods master the rich material created in the mathematical analysis by the finite differences and methods of linear algebra, and their numerous modifications connected with the equations of mathematical physics. In parallel, many specialists developed the notion of algorithms itself (Kolmogorov, Uspensky), and elucidate its universality, e.g., the independency from the concrete realization (thesis of Turing, Church, Post and Markov-young, which is often called as Church thesis). The theory of algorithm complexity began to develop as well (Cook and others).

However, at this stage the physical applications fit in this narrowly understood sense of “computational methods”. Tihonov and Samarsky ritually fixed this as the known triad: “model, algorithm, numerical solution (of the differential equation)”. Of course, this triad completely belongs to the classical mathematics, because its leading element – the model is the classical description of the system by means of differential equations. Algorithms here play the role of the technical service. The appearance of computational methods made possible to advantage to the borders of applicability of classical mathematics. This area includes all classical mechanics and electrodynamics, the thermodynamics, the theory of homogenous environments (aero and hydro dynamics, сыпучие среды, combined mass processes like nuclear reactions in the environment, etc), the theory of gravity and quantum mechanics, which in that times deals with quantum effects on the level of one or two particles. To nowadays the bulk of principal problems of this sort is already solved but those, which nature is purely quantum. We deal just with them. We agree that this area, which in nineteenth century belonged to physics (e.g. contained principal questions), was completely mastered by the computational methods. The further development of algorithm theory seemed predictable but at the end of twentieth century, the surprising intrigue appears concerning the quantum computer, which we discuss further. 

The main conclusion, which we must do from the development of algorithm theory and its applications, is that all reasonable procedures of theoretical physics are algorithmic.\footnote{My personal practice shows that the opposite arguments that sometimes appear in the scientific literature comes from the poor knowledge of their authors in algorithms. 
In the detailed analysis of concrete situations, which someone treats as non-algorithmic these arguments disappear as a mirage in desert. It does not concern the attempts to include to the area of theoretical physics such an object, as a human brain. In the dispute about this matter between A.A.Markov-young and A.N.Kolmogorov I take the side of the last one. In my opinion, we cannot include this object in the area of physics, at least while the constructivism yet does not reach the borders of biology.}

We can treat as simple the problems we spoke about earlier. We can define these problems as ones, for which quantum effects, needed for predictions, touch one or two particles. In the other words, these problems are physically simple. 

\subsection{Constructive mathematical analysis}

Constructivism represents the fundamental direction. This radically distinguishes constructivism from the computational methods, which bears the applicable sense. Constructive mathematics continued to develop in parallel and independently from the development of computational methods; this development made constructivism more and more important for theoretical physics. The next step was the creation of the constructive mathematical analysis, which authors were A.A.Markov young, and his pupil G.S.Tseitin.\footnote{The valuable deposit has been done by A.A.Muchnik, Y.Voscowakiks, B.S.Kushner and others.}

The sense of the constructive mathematical analysis is the refusal from the purely classical (and non-physical) notion of a real number. In the reality, we work not with real numbers themselves, but with their rational approximations, determined by some algorithm. These algorithms we treat as the constructive real numbers. It immediately gives us the non-solvability of the identification problem: it is impossible to determine algorithmically the equality of two given constructive numbers, though this is not an obstacle in the operations with such numbers. All real numbers, which can be determined by some reasonable way are constructive, for example, independently of their algebraic properties, for example, $\pi^e$, etc. The attempt to find an example of non-constructive number immediately requires non-algorithmic processes that have no physical sense.

We define a constructive function of the constructive real variable as the algorithm, which gives us the approximation $f_n$ of the function $f(x)$, given an approximation $x_n$ of the variable $x$. This approach to the definition of functions completely corresponds to physics because in the reality we never know the exact values of $x$ or $f(x)$, but only their approximations. Of course, if we fix some small threshold for the error $\e >0$, which we agree to treat as admissible, we can use the usual definitions. The principal difference appears only if we launch the error to zero. If it is possible to use the actual infinity in the classical mathematics, for example in the notion of the limit in the sense of Cauchy, in constructivism it is forbidden! Here we have no access to the exact values of $x$ and $f(x)$, but their approximations only. It leads to the surprising corollaries. Namely, in the constructive mathematical analysis there is the theorem of A.A.Markov the young and G.S.Tseitin, which says that 
\n

{\bf Any constructive function of a constructive real variable is continuous in each point of its domain}
\n 
 
It means the impossibility to use in constructivism non-continuous functions, like Heaviside function. Constructivism requires that the complete algorithm describe the real situation with all the details with involving all physical laws, which act in the considered area of configuration space. For example, considering an electron in the Coulomb potential on very small distances we must apply not electromagnetism only but also nuclear interactions. The practical application of constructivism presumes the realization of algorithms, which cannot work infinitely. 
All processes of sequential approximation must then finish. For example, we must cut potentials of Coulomb type in some manner that we presume implicitly to avoid the divergence of rows in quantum electrodynamics that is not a legal way in classical mathematics. In constructivism, this constriction is unavoidable, and the reasoning in QED obtains the legal status. 

Moreover, the constructive mathematical analysis completely preserves all analytical technique used in theoretical physics that makes legal all its results, e.g., all results of the modern physics remains unchanged in constructivism. Constructivism makes legal also a reasoning, which needs the complex substantiation in classical mathematics, and sometimes we cannot even reach such substantiation. These substantiations are not necessary in constructivism. Constructive physics does not require the exact values of magnitudes. Its aim is the different; we will speak about it further. 

\subsubsection{Constructive algebra}

A.A.Markov young started the constructive algebra by proving the theorem about the existence of a finite generated semi group with the algorithmically unsolvable problem of equality of words. It was the first train of the notion of the normal algorithm he introduced (the definition of algorithm equivalent to Turing machines). The connection between algebra and algorithms became explicit from this time, through the operations on words in the basic algebraic system – semi group, and it further developed on groups, rings, algebras and the other algebraic systems. \footnote{The further deposit to this direction was done by A.I.Maltsev, P.S.Novikov, S.I.Adian and others.} By a constructive algebraic system, we mean a system, which elements are encoded by the natural numbers, all operations are effective and the problem of equality of elements is solvable. This definition can be made more weak by various ways, for example we can consider the finitely generated systems for which the problem of equality not necessary solvable, etc. For the physical applications, it is important that matrix algebras, which constructive version uses constructive real numbers does not have the solvable equality problem, as constructive real numbers themselves, but support the analogy with its classical versions as well as in mathematical analysis. It means that all computations with matrix algebras remain valid in the constructive version of algebra. 

\subsection{Programming and its role in constructivism}

Programming appeared as the art of practical realization of algorithms. We here interest in natural problems, and I use the term modeling though the sense we acquire to it is more general than usual. By the modeling, we understand the description of the considered system evolution in time and the family of such evolution by a computer, such that these descriptions we formulate in any terms permitting the computer processing. For example, it can be the solution of the system of differential equations; it is the usual understanding of a modeling in the narrow sense. It may be also the direct modeling of the process where we directly represent its internal microscopic mechanisms as the operations or subroutines up to some known physical level (for example, to the level of a single molecule). 

The physical constructivism opens for the programming the new and exciting perspectives, because it obtains the much more high status than in the standard classical modeling. The point is that the model of quantum process requires the presence in it the specific subject called the observer. Quantum mechanics says nothing about how to distinguish this object from the others. We do not know, who or what can and who cannot be the observer. We presume only that the observer is the principally more complex system than the considered system. Moreover, it is impossible to include the observer into consideration as the object, which has some quantum state $|\Psi_{obs}\rangle$, since it must obey not quantum but classical laws. It means that in the entangling with it the state of considered object the collapse of the common wave function happens that is the act of observation. It means that there is not purely quantum mechanics, because for it something obeying classical laws must exist.  

This strange feature of quantum theory always involves the deserved admiration at this science. However, the preserving of this feature would make our main aim inaccessible. We must have such a model, which contains all the possibilities of scalability to the complex systems. It means that we have to include to our model also the mysterious phenomenon known as the collapse of the wave function. Our model must work in principle without any observer, e.g., the observations (that are measurements) must happen in it naturally, in its ordinary evolution. This requirement is very serious and we must explain it in more details. 

The uncertainty of the status of observer in quantum theory has the deep source. This source is the existence of principally new phenomena, which appears only in complex systems but not in simple. The main of such phenomena is the quantum entanglement. The other, not known yet phenomena are also possible, which concern the growing complexity of the considered system. Just the existence of such phenomena is the reason of division of the natural knowledge to the different sciences. We must take into account it in the realization of our project. This is why we cannot ignore the traditional structure of quantum theory. However, it turns that in the mathematical constructivism there is already the certain place for the observer! This is the place of the oracle in computations. 

Let we be given an integer function $F(x)$ of the unknown nature (for example, it can be non-computable, e.g., have no algorithm computing it). What means the computation with the oracle $F$? We imagine that the algorithm, accordingly which we perform computations, can in some instants query the oracle with the question: “what is the value of the function $F$ on the number $x$, which contains in the certain register of the memory?” After query, the algorithm temporarily stops the computation and waits the reply from the oracle. After the receiving of the reply, it puts it to the other certain register and continues the computation. Then it can repeat the query on the other word $x$ etc. It is the computation with the oracle. Thinking a little, we guess that the user of the program plays the role of the oracle. We then conclude that the user of the simulating program plays the role of the observer of the quantum system. The roles in the modeling are thus completely determined and in principal, we can begin to build programs. 

We must make the important notion about the role of the simulating programs in constructivism. The known fact is that it is impossible to predict the work of algorithm even on one-step ahead. There is the single one way to learn what an algorithm does: to launch it and observe its work step by step. It radically distinguishes an algorithm from its particular case – a formula, where it is typically possible to predict the behavior of the function it determines even to the infinity. This difference radically changes the criterion of success of a theory and the form of a theory in constructivism. We cannot have as a criterion the exact computation of some physical magnitude found in experiments that is the tradition of the usual physics. The criterion of validity of a physical theory in constructivism is the rightness of the dynamical scenario built by the modeling algorithm. 

A physical theory in constructivism is thus a heuristic for the building simulating algorithm, and the main object is the dynamical scenario. A theory is valid if this algorithm gives the right video film of the considered process. This is the success criterion in constructive physics. The exactness of the estimations of separate magnitudes is important only it helps to build the right algorithm, but not more. The exactness of estimations plays then the secondary role. 

The new role of programming is evident from this. In constructivism, it is the working tool of the theory. Its role is the same as the role of analytical computations in the traditional quantum physics. It brings the new requirements to programs, which constructivism development requires. These programs must be universal, not narrowly specialized. They must allow the simple introduction of the new particles and interactions in the considered system and permit to work simultaneously the different groups of specialists. Such programming seriously differs from its traditional treatment as the technology of the realization of complete mathematical solutions, and it requires the special attention. 

\subsection{What systems are complex}

We now explain what we will mean by a complex system. It is impossible to predict the evolution of such system by systems of differential equations or something of this sort. This definition is not satisfactory because it is non-physical. The impossibility to describe by systems of differential equations appears as the random character of the behavior of the system, even in the case when its usual behavior admits such description but there are singular points in which the evolution acquires the random character. For example, the behavior of the media can be in the critical dependence of a few molecules or atoms. Here we distinguish two cases. In the first case to get the complete description, we only need to apply classical mechanics to these atoms. In the second case, the complete description requires the application of quantum mechanics. It happens if the mechanical movements in the macroscopic volumes depend on chemical processes that involve a few molecules. We have no experimental evidences that these two types of complex systems are independent from each other. Moreover, we can certainly state the opposite. Namely, the source of any randomness in the Nature is some quantum phenomenon.  

One could object, on the example of the collisions of billiard balls, which in the case of large numbers of collisions give the completely chaotic picture, whereas the collisions are classical. The point is that the classical description is effective for a few sequential collisions only. If the sequential collisions are more than ten (and the number of the degrees of freedom is more than one – otherwise there is no place for the randomness), the classical description leads to the complete chaos. The point is that the error in targeting grows exponentially with the time that requires the corresponding increasing of accuracy, and it quickly leads to the necessity to consider the separate molecules on the surfaces of balls and their bounds inside of the ball, and the problems goes out of the framework of classical physics. 

We conclude that the systems representing the biggest difficulties for the simulation are quantum systems. The computational complexity then always has the physical basement. It is impossible to overcome quantum nature of complex systems. The theory of complex systems is thus the theory of quantum systems with many particles. 

\subsection{Why physics needs the constructivism}

We thus come to the main question about the connection of constructivism with physics. We formulate the conclusion definitely: to spread theoretical physics to complex systems physics must become constructive. This statement of question concerns the great project of quantum computer and it has the certain sense. 

\subsubsection{How quantum computer appeared}

We can really call the project of quantum computer (QC) the great because it decisively violates the usual order in the relation between physics and mathematics. In order to understand how this project lead us to the necessity of the constructive modification of quantum theory we need its detailed consideration. The source of QC is the same idea of Hilbert about the mathematization of the natural knowledge, and about the spreading of mathematical methods to complex systems. Since the way of axiomatization is closed, we have only one way: the simulation, or the modeling. The question is only what algorithms are appropriate for this. The building of the simulating algorithms is very serious thing, which we call the heuristic. The mathematical apparatus determines it. The traditional apparatus of quantum mechanics is classical mathematics, namely, the theory of Hilbert spaces. Here the state vector in the space of the dimensionality $exp(m\ n)$ determines the state of the system, where $n$ is the number of real particles in our system, $m$ is the number of grains in the configuration space accessible for the simulation, so that $m=1/g$, $g$ is the spatial resolution grain. The operator of evolution without measurements represents the unitary operator $U_t$, which depends on the time in this space. To simulate the evolution in this mathematical terms we need to work somehow with the matrix of exponential dimensionality that is impossible yet for three quantum particles (the case of two particles is peculiar; in quantum mechanics it is reducible to the case of two independent particles, as well as in classical mechanics). 

The first who recognized the fundamental obstacle was R.Feynman. He proposed to use for the simulation not the classical but the quantum computer. Quantum computer is the binary device on quantum bits, operations on which imitate the evolution of the real quantum system. This was the revolutionary idea violating the formed status-quo between physics and mathematics. At first, it turns that the notion of fast algorithms depends on physics. At second, algorithms themselves obtain the new power over physics, and it will never be the former science. The last thesis appeared not immediately, but after about ten years of rapid development of the mathematical quantum algorithm theory. 

At the first stage of the development of the quantum computing, it became clear that Feynman idea is blameless from the viewpoint of the standard formalism, which he possessed. Really, Wiesner and Zalka showed that it is possible to simulate the evolution of the abstract state vector in the space of exponential dimension by the quantum computer, factually operating with the quantity of qubits proportional to the volume of the system at hand. The time of such simulation we can make as close as desired to the real time of the modeled system that requires the simple change of the constant. 
Moreover, it turned that the quantum computer is able to solve also purely mathematical problems, for example to find the expansion to prime integers in the polynomial time (Shor) and the search problem in the time $\sqrt{N}$ where $N$ is the time of the classical search (Grover). Many other results appeared then formed the ideology of abstract quantum computing. For example, it turned that the bulk of classical computational tasks cannot be sped up on quantum computer at all. This underlines the peculiar character of quantum computations. There are not the usual computations in the classical sense. There are the special information procedures, which in some cases make possible to obtain the result much faster than any computation on a classical computer. Just this feature violates the usual non-formal bounders between physics and mathematics. Quantum computational procedures factually use some peculiar resource, named the entanglement, which does not exist in classical physics and in classical computations. 

\subsection{Quantum computer as the model object of the new physics}

If a quantum computer was built in its scalable version (versions with a few qubits already work, but this is the big difference), this course would never appear. The experimental work at its creating was the serious and sufficiently long that we can make some conclusions. Here the success would mean that the entanglement is the unlimited resource, from which we can extract the information factually without the deep analysis of its sense. The algorithm of Zalka and Wiesner makes possible to simulate complex chemical and biological systems and use all the advantages that it gives, without the deep analysis of the nature of the phenomena accompanying their evolutions. 

However, the different has happened. Quantum computer project slipped and buried hopes to conquer complex systems by means of the standard quantum theory! Formally speaking, quantum theory has not pass the test for effectiveness in the area of many body systems because experiments show that the role of decoherence substantially exceeds the effect of abstract quantum computing. For complex systems, the contra intuitive and non-mathematical part of quantum theory turns more valuable than the exact mathematical part, connected with quantum computations. We could predict this. One of the first who understood it was K.A.Valiev, who started the work in the quantum computer physics in our country. What conclusion can we make from this? Could we doubt in quantum theory itself, despite of that just it led us to the understanding of the real problems of the Nature? We could not do it even if it would not such a triumph of this science in the structure of atoms and molecules, optics and all the other happened in physics in 20 century. To make the right conclusion we should return to the thesis about the leading role of the mathematical formalism, which we discussed earlier. The conclusion can be only one. The classical mathematics was wrecked. Its pretentions to the conquest of the world of complex systems turn vain. The applications of algorithms in the traditional methods, giving the effect for classical systems turned out too week in the principally new situation when we study quantum physics of many body systems. 

All physical applications of the theory of differential equations have the certain sphere of applicability connected with the existence of the elementary quantity (grain) of each magnitude. The derivations of differential equations (oscillations, heat transfer, etc.) themselves rest on the application of the main law of interaction, which is limited by this grain. If $f$ is a quantity with some physical sense, it has the minimal nonzero value $\Delta f$. We consider quantum theory from this viewpoint. The wave function $|\Psi\rangle$ of any system must be then grained, e.g. it has the grain $\delta_\Psi$. We now consider the wave function of $n$ particles of the form $|\Psi\rangle=\sum\limits_{j=0}^{N-1}\la_j|\j\rangle$ where $N=O(exp(n))$ is the total number of basic states of the considered system. The typical value of $|\la_j|^2$ then equals $N^{-1}$. The detection of this magnitude requires the time of the order $N$, and it is impossible to estimate it practically already for the relatively simple systems of a few particles, let alone the practically interesting case of bio molecules. It means that we cannot in principal guarantee, that quantum states we create are really that quantum theory predicts. Let us suppose that the nonzero value $\delta_\Psi$ really exist and is sufficiently small to make no influence to quantum theory predictions for one or two particles (the case of two particles is reducible to the case of one particle). If we then increase the consider system, the typical value of amplitudes decreases exponentially, and we reach the threshold $\delta_\Psi$ very soon. The behavior of slightly more complex system will then differ substantially from the solution of Shredinger equation, which the experimenter treats as decoherence. We can show (see \cite{Oz}) that the existence of the amplitude grain $\delta_\Psi$ immediately gives Born rule for the computing of quantum probability. This is the constructive approach to quantum theory. Its specifications give us the mechanisms of modeling of the evolutions, which we call constructive heuristics. 

How can we help to build a quantum computer? We must study its physics. The physics of quantum computer is the key to its practical building. We do not know to what extend we can scale it, e.g., how much is the value $\delta_\Psi$ in the reality. Quantum computing seems now possible up to few qubits, may be up to tens qubits. However, where the border lies behind which decoherence makes the scalability impossible? I have not answer to this question. Do we need to develop quantum computations? Certainly, we need. We must understand how to realize basic principle of quantum computing on its real models with decoherence. It helps us to recognize its real borders and its nature; in the other words, we then hope to understand where the new physics of complex systems begins, which I spoke about earlier. 

However, we now cannot limit our efforts by the movement in one direction only. Classical mathematics discredited itself just in the area, where the support of mathematics is most important: in the theory of complex systems. There is no exact description of decoherence in the standard apparatus, whereas this phenomenon plays the key role in the physics of quantum computer. What solution of this problem proposes the constructivism? The constructive approach binds us to build algorithmic models of quantum evolutions on classical computers. Decoherence we must then treat as the unavoidable constriction of complex quantum states that happens due to the deficiency of the classical memory, which rapidly increases when the considered system grows.\footnote{Of course, the rest on classical computers does not mean the refusal from the usage of the limited quantum processors, for example, for the simulation of entanglement.} The building of such algorithms and programs requires the special heuristic, which we discuss further. 

We thus have to rest on the constructive mathematics. This presumes the serious change of the standard criterions and views. The main in this unavoidable matter is to get into the way of pluralism and take it as the native property of things. This side of constructivism is determinant for the success. Further, we see how to make quantum theory constructive. 

\newpage

\section{Lecture. History of collective behavior}

We saw that the main criterion in constructivism is the rightness of the video film about the considered process. This criterion is valid only in the case when the algorithm creating the video film is the general, e.g., it can process not the single but the wide range of systems. The modeling programs then must be universal. The universality in the programming requires the using of the instrument for developers that is the programming complex, which makes possible to elaborate the simulation programs for the particular cases. Speaking about the simulating algorithm, we will thus mean just this instrument for developers. Our algorithm must rest on the universal methods and notions and its structure must be completely transparent. The user must have the possibility to change the input data and check its work, etc.; we must guarantee the complete verifiability of the results. The structure of algorithm must be then universal and rest on the certain modification of the standard quantum theory. 

This modification we will describe in the next lecture. It is not the interpretation of quantum mechanics. It is the constructive truncation of quantum theory. It means that we reduce the formal capabilities of the Copenhagen quantum mechanics that makes it coordinated with the constructivism. We call such modification the heuristic. The different heuristics are possible depending on what lays in its basement. For example, one can formulate the heuristic so that it applies classical mechanics everywhere but the processes, which involve electrons, and for them it applies the wave functions. This way could give the initial advance to the chemistry, because this approach is typical in quantum chemistry. Nevertheless, this simple approach is not appropriate for our aims because it does not satisfy the property of universality. 

The method of collective behavior rests on the representation of one quantum particle in the form of the ensemble of point wise particles. This is not the new idea. The most known and successful allied approach is the formulation of quantum mechanics in the form of Feynman path integrals. 

\subsection{Feynman path integrals}

The remarkable principle of quantum mechanics is the principle of correspondence, according to which any physical magnitude corresponds the hermitian operator acting in the space of states of the system, where this correspondence ensures the passage from the quantum mechanics to classical for the value of action large in comparison with Plank constant. There is no quantum physics without classical in the strong sense of the word. There is not purely quantum theory; it must contain classical physics because the measurement procedure is impossible without the measuring device obeying classical laws. 

This impossibility to separate quantum physics from classical has the other sense concerning heuristics. The heuristic in the standard quantum theory as well as in the classical physics is non-formalized system of notions and agreements, which determines how we must apply laws and formulas to obtain the result having the physical sense\footnote{The analogy with a juridical system is pertinent where besides laws there are the procedures of their applications. It can be sub legislative acts, or precedents, or something else. It is important that this addition is unavoidable because laws will not work without it.}. The classical heuristic stands behind all advantages of quantum theory including the electrodynamics. This heuristic rests on the notion of a point wise particle. 

 The simplest form of systems is those contain only one particle or is reducible to one particle in the sense that the approximation of their dynamics by the simple combination of one-particle systems is satisfactory. The example is a particle in some potential. A real potential\footnote{We are always speaking about the electromagnetic potential. However, all we say about the quantizing is right for the other potentials including the nuclear and gravitation. For the last, the question is only how to divide it correctly to quanta.} is the sum of deposits from quanta of this potential - photons. Speaking about a particle in the potential, we mean the approximation resting on the following agreements:
a) there is the large total number of photons, and b) the emission of one photon does not cause the change of the source. We consider two interacting particles with coordinates $r_1$ and $r_2$. If we ignore quanta of the field again, we could introduce the new coordinates $R=(r_1+r_2)/2,\ r=r_1-r_2$ that reduces our system to two independent particles, e.g., to the simple combination of one particle systems. For three particles, this trick is impossible and the case of three particles represents the kind of a touchstone for the checking of various hypotheses in quantum informatics. 

In this paragraph, we consider systems reducible to one particle. For such systems in quantum case the classical heuristic is valid, which allows the reformulation of all conclusions of quantum physics on the language close to the classical (we call this language quasi classical). This heuristic allows the using of such terms as “trajectory", "the movement of particles along the trajectory", "deposit of the trajectory" etc., despite of that the formalism of wave functions contains no trajectories and no deposits. This powerful heuristic stands behind all advantages of quantum theory achieved to nowadays. Moreover, the constructive physics requires this heuristic because we have no alternative. R.Feynman did the first step on this way, and formulated quantum mechanics on the language of the so-called path integrals. 

The idea of this language is easy. Let us consider the flight of the particle from the point 1 to the point 2. It can fulfill this flight along the different trajectories. We accept that there is some algorithm determining for a given path $\g$ the number $A(\g )$, and for the different such numbers $A_1,A_2,\ldots, A_h$ their result $K(A_1,A_2,\ldots,A_h)$, which is in turn the number. The more the module $|K|$ is, the more probable our particle will occur in the point 2 provided it was initially in the point 1. This idea contains no numbers but only evaluations. However, it points to the algorithmic scheme answering roughly to the question where will pass the particle from the point 1. We conclude that a quantum particle factually represents not one point travelling in the space, but the set of points, each of which has its own path. These points are equitable in the sense that each of them equally represents the initial particle. It gives us the right to call them samples of the particle. 

A quantum particle looks like a galaxy, which stars are its samples, though this analogy requires some efforts in the description of interactions. 

It influence to the mixed consideration of systems by Born - Oppenheimer method. It considers atomic nuclei as classical, and electrons as quantum objects. 

In view of the above mentioned I call such a set of samples of the same real quantum particle a swarm that underlines its unusual character, where each sample represents the whole particle. 
It brings the question: what forces samples to hold together or what happens if some sample flies to far from the main part of the swarm. The answer presumes that we treat samples as non-erasable and the existence of some mechanism of the returning to the swarm distant samples. We consider this question further. 

It would be logically to treat each sample as non-erasable: that is to ascribe to each sample its own history. We return to this thesis further. Now we only treat samples as the auxiliary tool for the description of the wave function $\Psi$, and will redefine them anew in the short time frame $\delta t$, on the basement of the wave function $\Psi(t)$. This swarm we call wave swarm because {\bf its samples have only the history limited by the duration $\delta t$}. Our scheme then looks as the iteration of three main steps: 
\begin{itemize}
\item computing of the wave function from the states of all samples of the wave swarm, 
\item new definition of samples in the wave swarm,
\item free flight of samples and change of their amplitudes.
\end{itemize}

What is exactly the probability of the passage $1\ar 2$ ? This question concerns the interaction between samples of the swarm. The right answer is known: the probability equals the squared module of the value $K$, and we will establish why it is so. We are not bound by the necessity to consider the history of each sample and treat them as non-erasable; we accept that each sample is the carrier of some special number corresponding to it, and call this number its amplitude. To specify the details of this scheme we have to make more exact the meaning of the terms $K(A_1,A_2,\ldots,A_h)$ and $A(\g )$. The first term we define simply as the sum 
$$
K(A_1,A_2,\ldots,A_h)=\sum\limits_{j=1}^hA_j
$$
The second as 
$$
A(\g )=\frac{1}{A}\exp (\frac{i}{h}S[\g ]) 
$$
where $A$ is some constant and $S[\g ]$ is the action along the path $\g$, defined as  
$$
S[\g ]= \int\limits_{t_1}^{t_2}L(x_t,x,t)dt
$$
where Lagrangian of the considered system is $L=E_{kin}-E_{pot}$ the difference between kinetic and potential energy.

The function $K$ is thus complex valued and we can express it by the formula:

\begin{equation}
K(2,1) = \frac{1}{A}\int_{T(2,1)} \exp (\frac{i}{h}S[\g ]) D\g
\label{K}
\end{equation}
where the integration on $D\g$ means the summing on the set of all paths $T(2,1)$, going from the point 1 to the point 2. The wave function in the moment $t_2$ we can express through the wave function in the moment $t_1$ by the formula 
\begin{equation}
\Psi(x_2,t_2)=\int_{x_1}K(x_2,t_2;x_1,t_1)\Psi(x_1,t_1)dx_1
\label{ps}
\end{equation}
for any $t_1, t_2, x_2$. Here the point 1 has the coordinates $x_1,t_1$, the point 2 - the coordinates $x_2,t_2$. By the way, the formula \ref{ps} reflects the fact that the wave function is the value, which squared module is the probability density to find the particle in this point. This value equals $K$ provided the particle initially was in the point 1 (in the right side stands the action of delta- function on the wave function that immediately gives the desired). The formula \ref{K} formalizes the rule of computation of $K$, which we gave earlier.

Resuming the previously mentioned, we can define all three items in the evolution of the wave swarm. The computation of the wave function in the given point is the summing of amplitudes of all samples occurring in some cube with the center in this point. The redefinition of samples goes as follows. We divide the value of the wave function in the given cube to many peaces, e.g., represent it as $n\a$, where the natural $n$ is large, and create $n$ new samples. We ascribe to each of these samples the random speed from the uniform distribution on the large cube. At last, the free flight of samples goes accordingly to Galileo law, when all amplitudes are multiplied to $e^{-\frac{i\Delta S}{h}}$, where $\Delta S$ is the action of the sample along the straight line when the time $\delta t$ of its flight is fixed. To simplify the computations we agree that $\a$ is the same for all points and $n=n(x)$ proportional to $|\Psi (x)|$ in the given point $x$. 

The formalism of path integrals is equivalent to the standard. For this we follow the \cite{FH}, and find the value of the wave function defined accordingly to \ref{ps}, in the next time instant. We need the explicit representation of the kernel $K$. We suppose that for small values of the period $\e$ there is only one straight path of the form $1\ar 2$, and the integration in \ref{K} turns to one summand only. This trick requires the substantiation from the view point of standard mathematics, whereas in the constructivism it is legal, because if we limit the grain of spatial resolution by some threshold all trajectories become broken lines, which sections are straight lines and we obtain the desired. Analogously, it is possible to assume that the action equal the product of $\e$ and Lagrangian taken in some intermediate point. We then have
\begin{equation}
\Psi(x,t+\e )=\frac{1}{A}\int\limits_R\exp(\frac{i\e}{h}L(\frac{x-y}{\e},\frac{x+y}{2}))\Psi(y,t)dy
\end{equation}
Substitutions $L=mx_t^2/2-V(x,t)$ and $y=x+\delta$ give 
\begin{equation}
\Psi(x,t+\e )=\int\limits_R\frac{1}{A}\exp(\frac{im\delta^2}{2h\e})\exp(-\frac{i\e }{h}V(x+\delta /2,t))\Psi(x+\delta ,t)d\delta
\end{equation}
We see that the main deposit comes from the $\delta$ of the order $\sqrt{\e h/m}$. If we expand $\Psi$ to degrees of $\e$, keeping the summands of the first order, we must keep the summand of the second order of $\delta$. It gives with this accuracy
\begin{equation}
\Psi(x,t)+\e\frac{\partial\Psi}{\partial t}=\int\limits_R\frac{1}{A}e^{im\delta^2/2h\e}[1-\frac{i\e}{h}V(x,t)][\Psi(x,t)+\delta\frac{\partial\Psi}{\partial x}+\frac{\delta^2}{2}\frac{\partial^2\Psi}{\partial x^2}]d\delta ,
\end{equation}
and in our approximation with the equation 
\begin{equation}
\int\limits_R e^{im\delta^2/2h\e}d\delta=\left(\frac{2\pi ih\e}{m}\right)^{1/2}
\end{equation}
we have 
\begin{equation}
A=\left(\frac{2\pi ih\e}{m}\right)^{1/2}.
\end{equation}

Now we apply the known integral 
\begin{equation}
\int\limits_R\frac{1}{A} e^{im\delta^2/2h\e}\delta^2=\frac{ih\e}{m}
\end{equation}
and obtain 
\begin{equation}
\Psi+\e\frac{\partial\Psi}{\partial t}=\psi-\frac{i\e}{h}V\Psi -\frac{h\e}{2im}\frac{\partial^2 \Psi}{\partial x^2}
\end{equation}
which immediately gives Shredinger equation. 

The method of path integrals thus represents the version of quantum mechanics. Its practical implementation presumes the ingenious computations with integrals and algebra, and it gives the new insight into quantum theory in comparison with Shredinger equation. For example, it is possible to find the kernel of particle in the different fields (see \cite{FH}), for the free particle, it has the form 
\begin{equation}
K(x,t,0,0)=\left(\frac{2\pi iht}{m}\right)^{-1/2}e^{imx^2/2ht}
\label{ker_free}
\end{equation}
which shows the character of the movement of its samples to the points $x$ with the speeds $v=x/t$ that corresponds to the classical picture of the flight of pieces after the "explosion" of the particle initially located in the center of the coordinate system. 

Path integrals allow the deduction of the uncertainty relation of the type "coordinate - impulse". It follows from the consideration of the passage of the particle trough the narrow slit of the weight $2b$. Slightly clumsy calculations of the kernel of the passed particle (see \cite{FH}, pages 63-64) show that after the passage through the slit the support of the wave function obtains the widening $ht/mb$, which means the obtaining of the additional uncertainty in the impulse as $h/b$, that gives the uncertainty principle of Born- Heisenberg:
\begin{equation}
\delta p\delta x=2h.
\end{equation}

In the formalism of path integrals the notion of sample plays the secondary role since each sample of the real particle preserves its history in the short time frame $\delta t$ only\footnote{There is no explicit notion of sample in the original text \cite{FH} at all.}. The main role belongs to the wave function, which factually determines how many and what samples we must create in each spatial cube. However, just the free movement of samples is the step of evolution creating the unitary dynamics. The secondary character of a sample in the wave swarm follows from that the density of such swarm is proportional to the module of the wave function, not to its square. In the other words, the density here if not the density of probability of finding the particle in this point. The brevity of the sample history leads to the absence of the dynamical characteristic of the sample themselves. The mass of sample, which in this approach is equal to the mass of the real particle occurs only in the evolution of its amplitude through the action. Despite of the ephemerality of the wave swarm samples their application gives much. In particular, we can by means of path integrals formulate the criterion for what dynamics must we apply for this particle: quantum or classical. 

Indeed, let us consider the passage from the point 1 to the point 2 along two paths: $\g_{cl}$ the path, which is the solution of the classical dynamics equations and $\g_{noncl}$, which is some other path. Without loss in generality we can accept that the samples preserve their history travelling from 1 to 2, e.g., that these point are sufficiently close to each other. We compare two deposits to the wave function: 1) the deposit from amplitudes of samples flying along trajectories close to $\g_{cl}$, and the deposit of samples flying along paths close to $\g_{noncl}$. We denote them by $K_{cl}$ and $K_{noncl}$ correspondingly. Let $S_{cl}$ be the action along the classical path. We suppose that  
\begin{equation}
|S_{cl}|\gg h.
\label{crit}
\end{equation} 
We can suppose that the change of action on the order of value equals to the action itself if the path have the general form. Since the classical path $\g_{cl}$ is the solution of the equation $\frac{\delta S[\g ]}{\delta \g}=0$ (the principle of least action) the change of the phase on all paths close to  $\g_{cl}$ is small. The change of the phase on paths close to $\g_{noncl}$ is large because of the inequality \ref{crit}, because here the equation $\frac{\delta S[\g ]}{\delta \g}=0$ does not take place. It means that the deposit $K_{cl}$ on absolute value is much more than the deposit $K_{noncl}$, due to that in the first deposit we sum the values with the approximately equal phases, whereas in the second the phases are different. We thus obtain that all the amplitude concentrates along the classical path of the particle, e.g., it behaves as a classical particle. Whereas if \ref{crit} is not true, the situation changes, and the deposit $K_{noncl}$ can compete with $K_{cl}$, that is the real particle will not move only along the classical path, and will reveal quantum properties, for example, interfere in the passage through two slits, etc. We then can accept that the inequality \ref{crit} is the criterion of the applicability of classical mechanics. We also note that the smallness of the action can follow from the lightweight of the particle, small speed, or small period. Even massive particles moving slowly in the short time frames reveal quantum properties. In the practical description the period $\delta t$ we take such that it gives the pithy general picture (not only the isolated particle dynamics). Hence, for example, electrons typically reveals not as point wise objects, but as the wave functions, whereas nuclei are classical objects. This is Born -Oppenheimer model. It is convenient for the cases as atomic physics, where the subject is many electron states, or molecular dynamics, where the stretching and rolling of molecules are in focus. In opposite, this model is not always appropriate for chemical reactions where the quantum character of the movements of nuclei plays the key role. In the same reaction often is necessary to have the classical and the quantum consideration of reagents. Born -Oppenheimer model is convenient mainly due to its simplicity resting on the large (in about 2000 times difference between the masses of electrons and the proton. 

\subsection{Monte Carlo method}

For the search of the ground state (the eigen state with the minimal energy) the most convenient is the method of Monte Carlo, which we can call the method of static diffusion. This method represents a quantum particle as the swarm of its samples so that its density equals to the module of the wave function: 
$\rho =|\Psi_0|$. It shows the narrowness of this method because it cannot serve as the probability model of quantum dynamics where the density must equal the square of the wave function module. The deep source of this difference is the static character of Monte Carlo method, which is not appropriate for the modeling of dynamics. The dynamics requires the description of excited states, not only the ground state. However, Monte Carlo method is simple in usage and we can treat it as the good starting point for the creating of the real dynamical model. 

We consider the following process of the swarm evolution. Each sample with some small probability $p$ jumps to $\Delta x$ along one of coordinate axes to the positive or negative direction with the equal probability $p$. With the probability $1-6p$ a sample stands at place. Let we be given a scalar function $V$ on the configuration space. We also accept the following rule of creation and annihilation of samples. Let the following process happen for each sample with the probability proportional to $V$. If $V<0$, then this sample generates the new sample located in the same point, if $V>0$, then this sample eliminates some other sample located on the distance lesser than $\Delta x$. We call this process the static diffusion. Samples have no speeds here and only their coordinates participate. 

Now we make in Shredinger equation 
$$
ih\frac{\partial\Psi}{\partial t}=-\frac{h^2}{2m}\Delta\Psi +V\Psi
$$
the formal replacement of the variables $t=-i\tau$. The equation then acquires the form of diffusion equation:
$$
h\frac{\partial\Psi}{\partial t}=\frac{h^2}{2m}\Delta\Psi -V\Psi
$$
which just describe the stationary diffusion process. If we put the energy levels into order of the growing energy values $E_n$, then the evolution of the state vector expanded to eigenvectors $\phi_n$ will be:
$$
\Psi=\sum\limits_n \la_n e^{-\frac{i}{h}E_nt}\phi_n .
$$
The expansion of the diffusion equation then obtains the form
$$
\Psi(t)=\sum\limits_n \la_n e^{-\frac{1}{h}E_nt}\phi_n .
$$
We see that for $t\ar\infty$ the diffusion process converges to the ground state, because the rapidly decreasing exponentials suppress all other states. It is known that the ground state contains no phase differences, e.g., it coincides with its module. Hence, to find it, we have to determine the initial swarm density distribution and launch the stationary diffusion process. It stabilizes on the distribution proportional to the ground state. Of course, we should tae care of the conservation of the total number of samples in some reasonable frameworks. We can easily guarantee it by the addition or elimination samples uniformly accordingly to the existing density. 

DMC method can be easily generalized to the case of $n$ particles. A sample here will be a cortege of $n$ samples, each from the swarm of the corresponding real particle, and the configuration space will be $R^{3n}$.

We note again that DMC method is aimed to the search of the ground state only, e.g., for the stationary modeling. To find the dynamical picture we need the method of dynamic diffusion, which we describe further.

\subsection{Bohm approach}

The method of Bohm uses the notion of pseudo potential.
We identify the square of the wave function module with the density of the swarm and its phase $\phi$ with the classical action. We then have $\Psi = \rho^{1/2}e^{i\phi /h}$, and $1/m\ grad\ \phi (\bar r)$ we can regard as the density of the swarm stream. Shredinger equation then becomes equivalent to such system of equations:\footnote{These equations derived Madelung. Bohm lied them in the basis of his interpretation of quantum theory; we can treat it, with Feynman path integrals the pre image of the collective behavior method, which we study further.}
$$
\begin{array}{lll}
&\frac{\partial\rho}{\partial t}+div\ (\rho/m\ grad\ \phi )&=0,\\
\frac{\partial\phi}{\partial t}+&\frac{1}{2m}(grad\ \phi )^2+V+V_1&=0,
\end{array}
$$
where the quantum pseudo potential $V_1 = \frac{h^2}{m}(\Delta\rho/\rho+(grad\ \rho )^2/\rho^2 )$ depends on the density of samples with the singularity in the initial point. These equations coincide with the equations of classical particles dynamics if we accept that $V_1$ has the form of some potential. 

\subsection{The drawbacks of analytic approach to collective behavior}

We sum up the tricks using the idea of collective behavior: the exact reformulation of quantum mechanics on the language of Feynman path integrals, Bohm approach and diffusion Monte Carlo method. They are different in aims but have the common idea of the representation of a quantum particle as the swarm of point wise particles. They all rest on the classical mechanics that determines the limit of their applicability. 

We begin to analyze them with the most valuable Feynman path integrals. In the proof of equivalence of this method with Shredinger equation, we use the explicit representation of trajectories in the form of straight-line segments on the small distances. It is natural for the constructive version of collective behavior, which determines the dynamics of the separate samples by the sequence of such steps: 
\begin{itemize}
\item free flight of all samples in the period $\Delta t$,
\item interaction of close samples resulting in the change of their speeds by some local rule.
\end{itemize}
This is the steps of constructive method of collective behavior, which we call the method of dynamic diffusion; we describe it in the next lecture. However, from the view point of classical mathematical analysis the supposition about the straight form of trajectories on the small periods cannot be proven and we must take it as an axiom. Moreover, the analysis of quantum trajectories by path integrals (see \cite{FH}) shows that the main deposit to the amplitude belongs to n0n-regular trajectories, that is the paths, which have the more differences in speed the less spatial grain we use for their representation. This property of the fast change of speed on quantum trajectories we can illustrate using the uncertainty relation “coordinate – impulse”: the less is the grain of spatial resolution, the bigger is the dispersion of speeds inside of each grain. It cause difficulty in the formal substantiation of path integral method in the classical mathematical analysis.\footnote{Difficulties in the substantiation of formal methods, which give the maximal effect in theoretical physics are typical. We can recollect generalized functions, asymptotic rows, etc. The typical reaction of physicists to these difficulties is the following: let mathematicians take it up; it is their job. This position is pragmatic but narrow sighted. The effectiveness of the usage of mathematical apparatus depends on the accuracy in the substantiation of used methods, as the secure driving depends on the satisfaction of traffic rules. The violation of these rules leads to the crashing results immediately when the usual situation: for example, in the area of many particle systems, containing quantum computers.}

Diffusion Monte Carlo method gives good approximations for wave functions of ground states. However, it is not applicable for the description of dynamics and is not thus completely constructive. Samples in DMC method have no speeds, only positions and it determines limitations of this method. 

Bohm approach contains the serious drawback from the constructive viewpoint. The mechanism of interaction between samples, which could give this potential, is unclear. Moreover, the singularity in the initial point means that in the areas with the small density some huge force acts on samples with the unpredictable direction that makes impossible to apply this approach to the computer simulation of quantum dynamics. \footnote{D.I.Blohintsev treated in \cite{Bl} the impossibility of the simple interpretation of linear property of Shredinger equation solutions in hydrodynamic terms as the main drawback of this way. I think that it is not a lack at all. It is impossible to overcome the contradiction between the stochastic nature of the wave function and linear properties of Shredinger equation in any ensemble interpretation of the wave function. However, it is not required for the building of effective simulating algorithms. Any method of matrix algebra has the genetic lack in the usage of computational resources; for the simulation of many particles, we unavoidable have to choose between some version of ensemble representation and quantum computer. This is why the sacrifice of the algebraic beauty would be reasonable and necessary here.}

The deep source of this difficulty lies in the statistical nature of the wave function. Its experimental finding always requires the large number of the repeated experiments and the linearity can reveal exclusively as the limit result of them if their total number goes to infinity. The refusal from the simple representation of linearity is not thus the serious drawback. Moreover, the effective modeling requires the refusal of something more. We need the simple mechanism of the interaction between samples in the dynamic diffusion swarm. It turns that for the obtaining of such mechanism we must refuse from the relatively beautiful systems of differential equations, as was written earlier. It is possible to write the system of differential equations only for the fixed spatial resolution. This is the price, which we must pay for the method of building of the effective algorithms, e.g., this is the price of the constructive modification of quantum theory! 

At last, the important deficiency of classical versions of the collective behavior that we have studied here is the lack of the individual history of samples. Factually, it is true also for the DMC method because we have to renormalize the wave function periodically, which means the addition or annihilation of samples. We further return to the discussion of the role of individual history of samples. Now we only note that the absence of the individual history deprives classical versions of the collective behavior the additional predicting force in the comparison with Shredinger equation itself. Using of the classical mathematical analysis as the main tool in theory makes impossible to us to go out of the framework of standard problems contained in traditional courses of quantum mechanics, let alone to study complex systems. 

\newpage
\section{Lecture. Dynamic diffusion swarm}

We now describe completely constructive form of collective behavior, which can serve as the heuristic of the modeling algorithms. 

   We call it the method of dynamical diffusion swarm. It differs from the previous attempts in that we set the aim not the obtaining of the differential equation but the mechanism of interaction giving quantum dynamics by the shortest way for the realization on a classical computer. 

   Why the dynamical diffusion swarm is better than the solution of Shredinger equation by one or the other method? In the solution of a differential equation, we factually use Riemann scheme of integration. Computations of the wave function go in all configuration space independently of how constructive the interference of amplitudes is. On the major part of the space, where the interference is destructive and the wave function factually equals zero we then have to spend the computational resource only to check it. The dynamical diffusion swarm, in contrast, realizes the more general, Lebesgue integration scheme. In this scheme, the diffusion dynamics will concentrate diffusing particle in the areas of the constructive interference that gives the great saving of the computational resource. This is the fundamental advantage of the diffusion dynamics. We will see that the cost of this advantage is the non-uniform intensity of the diffusion on the element of space. This intensity will depend on the grain $\d x$ of spatial resolution chosen in advance, in contrast with the usual diffusion, where there is not such dependence. 

   We proceed with the definitions. We call a swarm a finite set $S$ consisting of $n$ point wise classical particles of the same type, each of which  $s\in S$ possesses its coordinates and impulse $x(s),\ p(s)\in R^3$. The collective behavior method represents one quantum particle of the mass $M$ and the charge $Q$ by the swarm $S$, which samples $s\in S$ have the mass $m=M/n$ and the chargeд $q=Q/n$. Members of this swarm we call samples of the particle. We assume that the total number of samples $n$ is so huge that it can serve as the approximation of the continuous media. If we need to pass to higher and higher resolution there will be samples in each cube of the corresponding subdivision of the space. The dispersion of the speeds of samples will grow when the grain of spatial resolution decreases. It means that we will have the separate swarms for the different values of the grain.

   The methods of determination of particle depend on the specific problem, so that particles are not necessary elementary in the sense of theoretical physics. The definition of what to treat as a particle presumes the fixation of the typical length $\Delta X$ and time $\Delta T$, so that if the size of a particle is much less than $\Delta X$, then we treat it as a point wise, and the time period $\Delta T$ must not be less than the time of process we consider. We also assume that the typical medium speeds of movements are much less than some limit speed for the movement of real particles $c$.
For example, we can treat an atom as point wise particle in processes with $\Delta X > 10^{-8}m$ and $\Delta T>10^{-10}s$. If we make decrease the value of typical length and time, then for the right dynamical picture we need to consider the other set of elementary particles, for example, we can consider separately a nucleus and electrons inside of atoms. Fixing $\Delta X$ and $\Delta T$, we must determine smaller segments $\d x$, $\d t$, which represent elementary steps of the video film, though they are lesser than the typical lengths and times $\tilde\Delta X,\ \tilde\Delta T$ for the more fundamental processes than we regard. The gap between these values can be about $10^{-20}$ that makes the distinguishing possible). In the process with the fixed energy, the lengths and times depend on the mass of particles. The separation of particles by their masses in electrodynamics allows the consideration of only electrons, because the typical distances of nuclei movements will be $1800$ and more times lesser. We then can regard the chosen values $\Delta X$, $\Delta T$ as the relative size of an imaginary screen and the imaginary duration of the film, and $\d x$, $\d t$ as the screen resolution and the time of showing of one cadre in our video film. We choose $\d x$ and $\d t$ as large as possible that make our video film informative. After this choice we can determine, which particles are quantum and which are classical. For it we compare their typical actions $a=M(\Delta X)^2/\Delta T$ with Plank constant $h$. If $a<h$ the particle should be treated as quantum, otherwise as classical. We will see that in the method of collective behavior the passage from one type of consideration to the other means the change of swarm size, e.g., does not concern the introduction of the different types of dynamics. In view of the reserve we noted, and the choice of resolution we will be able in the preparation of the film decrease the values $\d x$ and $\d t$ in order to form the right image, for example, by the additional subdivision of this segments and substantiate the quality of the approximation to solutions of Shredinger equation. We assume that the space $R^3$ is divided to the equal cubes with the side $\d x$, and the time to equal periods $\d t$. 

We introduce some value of speed $c$, which we treat as the limit for the moving of samples. Segments of distance and time we choose such that $\d x \gg c\d t$. It guarantees that at each step of evolution the values of magnitudes obtained by the averaging on cubes with the side $\d x$ will vary slowly that is necessary for the asymptotic approximation. 

The density of swarm in the point $x$ is given by the expression 
\begin{equation}
\rho (r, t) = \frac{N(r,t)}{(\d x)^3},
\label{density}
\end{equation}
where $N(r,t)$ denotes the number of samples occurring in the instant $t$ in the same cube with the point $r$. For the comparison with the solutions of Shredinger equation we would launch in these definitions $\d x \ar 0$, which means that we consider not one swarm but the sequence of swarms with densities $\rho_n$ and increasing $n$. We will not do this in the order to simplify notations; instead we assume that it is always possible to split additionally the division of space to cubes so that $\d x$ will be lesser in the fixed frameworks. We will write $\rho(x)=|\Psi(x)|^2$, which means that  
\begin{equation}
\rho_n(x)\ar |\Psi(x)|^2 (n\ar \infty ),
\label{asympt}
\end{equation}
where the convergence will be uniform, without the special mentioning. This sequence of swarms, realizing the approximation to the density of the wave function – solution of Shredinger equation, we call the admissible approximation of quantum evolution. 
    
  Our main aim is to define the behavior of samples, which gives the admissible approximation to the quantum evolution.

The main requirement to the simulation of quantum dynamics via collective behavior is the following.
\begin{itemize}
\item The swarm dynamics simulates the quantum dynamics so that in each instant $t$ the quantum probability equals to the swarm density 
\begin{equation}
|\Psi(x,t)|^2=\rho(x,t)
\label{swarm}
\end{equation}
 in each point $x$ of the configuration space. 
\item Each sample has its own history, e.g., it preserves its individual identification number during all the simulation. The types of samples exactly correspond to the types of real physical particles. 
\item The behavior of each sample is completely determined by its own state, the state of all samples in its vicinity and some source of random numbers.
\end{itemize}

A swarm satisfying these conditions we call a quantum swarm for one quantum particle. 

We define the behavior of samples so that these conditions are satisfied. For this it would suffice to show that for any solution $\Psi(x,t)$ of Shredinger equation we can move samples only locally, e.g., to small distances and thus ensure the equality (\ref{swarm}) in each time instant. 

 The second rule means that we refuse from the complex numbers and at the same time want to include QED in our model in future. The last requires the locality of all interactions. The behavior of a sample is the rule determining the change of its state (impulse, momentum of impulse and the type) and the spatial position (spatial shift). As we know, the behavior cannot rest on classical mechanics. 

We define the quasi-classical behavior of samples called the dynamical diffusion mechanism. The swarm with this mechanism will satisfy our conditions. 

We accept that each sample in each instant can either stand at place, or move with the speed $c$ along on of the coordinate axes $OX,OY,OZ$. 

The reaction of exchange we call the following sequence of actions:
a) the choice of the pair of samples $\a, \b$, the distance between which is not greater than $\Delta x$, which speeds are mutually opposite: $v(\a )=-v(\b )$ and b) either the simultaneous replacement of their speeds to zero (if they are nonzero), or the acquirement by them the mutually opposite speeds of the module $c$ directed along one of coordinate axes. 

The exchange does not change neither the summed impulse of the swarm, nor the summed momentum of impulse provided $\Delta x$ is sufficiently small. 
We denote by $N(r)$ and $N_s(r)$ the set of samples in the cube with the point $r$ and the set of samples with zero speed inside it, by $N^+_x(r), N^+_y(r), N^+_z(r)$ we denote seta of samples from the cube with $r$ moving along the corresponding axes in the positive direction, and by the analogous symbols with the sign $-$ - in the opposite direction. By $|g|$ we denote the quantity of samples in a set $g$. We agree to denote the quantity of samples in a set by the same symbol, as this set, but with the replacement of $N$ by $n$. We call $r$ - stationary each subset $S\subseteq n(r)$, consisting of samples with nonzero speeds for which $\sum_{\a\in S}v(\a )=0$ and $S$ is the maximal with this property. The number $|S|$ of elements of $r$ - stationary set (which does not depend of its choice) we denote by $s(r)$. Let $d>0$ be the chosen constant so that the coefficient of diffusion is proportional to $d$, $V(r)$- is the scalar field proportional to the external potential energy with the constant coefficient of proportionality, $grad\ V(r)=(V_x(r), V_y(r), V_z(r))$.  

We will consider only non relativistic swarms, e.g., such that $n_s(r)/n(r)$ is close to $1$ for all $r$. It means that the bulk of samples in each cube have the zero speed. This requirement is incompatible with the point wise approximation \ref{asympt} of the exact wave function by swarms for external potentials of the Coulomb type $1/r$ because the mean speed near the initial point goes to infinity. To have the asymptotic convergence \ref{asympt} we must be able to choose the speed $c$ as large as we need for the regular swarm of the number $n$. In the reality $c$ cannot exceed the speed of light that establishes the natural limit to the accuracy of the swarm approximation of the solutions of Shredinger equation. 

We call the dynamical diffusion mechanism of the evolution the sequence of the following actions with the swarm. 

\begin{itemize}
\item 1) The sequence of random exchanges with the uniform distribution leading to the distribution of speeds with the property $s(r)/n_s(r)=d$ for each point $r$. If $n(r)$ is small, this equation must be true with the maximal accuracy. 
\item 2) The ascription of the speeds to some samples from $N_s(r)$, chosen randomly from the uniform distribution so that the signs of newly obtained speeds along each coordinate axes are the same and if $v_u(r)$ is the summed speed obtained by samples from the cube $r$ along the axes $u$, $u=x,y,z$, then for such $u$ the equation $v_u(r)m=-V_u(r)$ is satisfied with the maximal accuracy.
\item 3) The change of coordinates $r(\a )$ of each sample accordingly to the law of uniform movement:
$r_{new}(\a )=r(\a )+v(\a )\Delta t$.
\item 4) The converting of $V(r)$ accordingly to the new coordinates of particles.
\end{itemize}

We do not specify the method of converting of the potential energy. We can do it by the Coulomb law or by the diffusion mechanism proposed in the work \cite{Oz1}. 

The swarm with the dynamical diffusion mechanism of evolution we call the dynamical diffusion swarm (DDS). This swarm is irreducible to the ensemble of point wise particles with the classical interaction. The point 1) says about two things:
\begin{itemize}
\item there is the random force of repulsion or attraction, which preserves the summed impulse of the swarm (compare with \cite{FH}), and
\item the averaging of speeds of samples goes with the accuracy determined by $d$ (the less the value $d$ is, the more exact averaging we have.
\end{itemize}

For each time instant $t$ if $\Delta x$ is sufficiently small the density of swarm $\rho (r,t)$ for any point $r$ will not depend of the coordinate axes orientation.
Really, let $\d_1$ be such that $c\d t\ll\d_1 x \ll \d x$, and $v(r)$ denote the mean speed of samples in the point $r$, found by the averaging on the samples with coordinates $r_1:\ \| r-r_1\| < \d_1 x$. The total number of samples came in the unit of time from the vicinity of the point $r_1$ to the vicinity of the close point $r_2$ will be proportional to the scalar product $v(r_1) (r_2-r_1)/\| r_2-r_1\|^2$, which is independent from this orientation. 

We compare the defined dynamical diffusion swarm with the ensemble used in diffusion Monte Carlo method, which we call the stationary diffusion ensemble. The determining of the dynamical diffusion swarm state means the defining coordinates and speeds of its sample whereas only the density determines the state of stationary swarm because the mean speed of its sample is zero. Indeed at the step of evolution of stationary swarm for each its sample we choose with equal probabilities one of its shift: up, down, to right or to left, or it stays at place. This choice is independent from the initial speed of the sample and the resulting impulse is always zero. This model corresponds to the spreading of molecule of some substance in the media of the other substance< for example, of the color in the water. If we trace only the molecules of the color, the impulse will not be stationary: it passes to the molecules of the water. The stationary model of diffusion we can thus call the model with the friction.

In contrast to the stationary swarm, the dynamical diffusion swarm evolves with the conservation of impulse in each interaction of pairs of samples in the exchange. Not only its density, but also the summed impulse in each cube of the space then determines the state of the dynamical swarm. E.g., the pair of functions $\rho (\bar r, t),\ \bar p (\bar r, t)$ then determines the state of DDS.

However, the most important advantage of DDS in the comparison with the known ensemble methods is that each sample has its own history. It substantiates the name of sample. DDS allows the uniform consideration of a sample: from the classical and from the quantum viewpoints. To pass to the quantum consideration from the classical and vice versa we need only to replicate samples (or, conversely, to merge them). Here the swarm corresponding to the classical particle contains only one sample. We can also regulate the allocation of the computer memory in the simulation taking into account that the exactness of the quantum description grows with the total number of samples in the swarm. 

\subsection{Differential equations for dynamical diffusion swarm}

To prove the appropriateness of the diffusion dynamics for the approximation of Shredinger equation we need to pass to the differential equations with the functions of real variables. Here the difficulty is that such equations will depend on the elementary length $\d x$. For example, the intensity of the diffusion process is proportional to $(\d x)^{-3}$. It makes impossible to launch $\d x$ to zero as in mathematical analysis, when we use it to the classical physics. The fixed grain of spatial resolution $\d x$ determines the status of equations, which we will write. We must choose it such that the approximation of the density $\rho = |\Psi |^2$ of the wave function by the density of the diffusion swarm within $(\d x)^3$ is satisfactory for the considered process. After the fixation of $\d x$ we can consider the dynamical diffusion swarm of the corresponding intensity and the differential equations approximating its dynamics, which will turn to equivalent to Shredinger equation. 

The following pair of functions thus gives the state of the dynamical diffusion swarm 
\begin{equation}
\rho(t,\bar r),\ \bar p(t,\bar r),
\label{pair}
\end{equation}
where $\rho$ is the scalar function of the density of samples, $\bar p$ is the vector function equal to the summed impulse of samples in this point, which we define as $\lim\limits_{dx\ar 0}P(r,dx)/(dx)^3$, where $P$ is the sum of impulses of samples occurring in the cube around $r$ with the side $dx$. Here we assume that $dx$ can be done substantially lesser than the chosen value of the grain $\d x$, which defines coefficients of equations on  $\rho$ and $\bar p$. 

The dependence of the equation on the grain $\d x$ looks as follows. The summed impulse $\bar p(t,\bar r)$ will vary slowly when $\bar r$ varies to the values greater than $\d x$. But its derivative $\frac{\partial \bar p}{\partial t}$ is huge: of the order $1/(\d x)^3$, and varies quickly. E.g., the graph of function $\bar p(t,\bar r)$ will be smooth if we observe it with the big grain $\d x$, but if we increase the resolution by decreasing the grain $\d x$, we see that it looks like a saw with sharp teeth. The sharpness of teeth will be the more, the more is the resolution $1/\d x$, and is limited by the maximum speed of samples $c$ only (compare with \cite{FH}). 

We express this requirement in the form:

{\bf The intensity of the impulse exchange between the neighbor spatial areas must be much more than the intensity of sample exchange.}

This condition is important, we call it the non-relativistic approximation and write it as $v\ll c$. Factually, this requirement means the deviation from the classical physics in the interaction between samples of quantum particle, as the special character of impulse exchange. In the classical impulse exchange, we would obtain the oscillation equation with the second derivative on the time. The impulse exchange in the dynamical swarm has the peculiar character when samples standing at place simultaneously obtain mutually opposite impulses. 

In view of the isotropic property of the diffusion process the change of density $\rho(r,t)$ in the time and its second derivative has the following expression through the integration on the surface $S(r)$ of the sphere of radius $\d x$:
\begin{equation}
\begin{array}{lll}
&\frac{\partial\rho(r,t)}{\partial t} &=\frac{3}{4\pi (\d x)^3} \int\limits_{S(r)}\bar p(r,t)\bar n(\bar r_1)ds(r_1),\\
&\frac{\partial^2\rho(r,t)}{\partial t^2} &=\frac{3}{4\pi (\d x)^3} \int\limits_{S(r)}\frac{\partial\bar p(r,t)}{\partial t}  \bar n(\bar r_1)ds(r_1).
\end{array}
\label{density}
\end{equation}
 These formulas follow immediately from the definition of density of samples and are true for any mechanism of the movement. 

We deduce the law $\frac{\partial\bar p}{\partial t}\bar a$ of change of the summed impulse of the swarm in the small ball with the center in the point $\bar r$, caused by the movement of samples along the normal vector $\bar a$ to the surface of this ball of the unit length. 
The deposit to the summed impulse of the small ball consists of three values:
\begin{itemize}
\item Penetration of samples, obtained the speed in the exchanges through the small element of square. 
\item Penetration of samples, obtained the speed from the action of the potential $V$.
\item Penetration of samples preserved their speed, e.g., by the inertia.
\end{itemize}
It follows from the definition of the dynamical diffusion that these deposits are correspondingly,
$-I\ grad\ \rho\bar a$, $-\kappa\rho\ grad\ V\bar a$ and $g\rho\bar p\bar a$, where $I,\kappa,g$ is the intensity of the corresponding processes. By the appropriate choice of the unit system we can guarantee that $g=1$. The following formula expresses the dependence of the coefficients on the grain of spatial resolution, which is necessary for the approximation of Shredinger equation:
\begin{equation}
I=\frac{h^2}{2m^2(\d x)^3},\ \kappa = \frac{h}{m\d x}.
\label{intens}
\end{equation}

In view of the non-relativistic approximation we can reject the last summand, which is substantially lesser than the first two for the small $\d x$. It gives the formula:

\begin{equation}
\frac{\partial \bar p}{\partial t}\approx -I\ grad\ \rho-\kappa\rho\ grad\ V.
\end{equation}

We thus obtain the equation on the density of the diffusion swarm:
\begin{equation}
\frac{\partial^2\rho (r)}{\partial t^2}=-\int\limits_{S(r)}I\ grad\ \rho -\kappa\rho\ grad\ V)\bar n(r')\ dS(r'),
\label{swar}
\end{equation}
where coefficients $I,\kappa$ are given by formulas \ref{intens}. 

We prove that the quantum swarm satisfies the equation \ref{swar} that means the approximation of quantum dynamics by the swarm evolution. 

\subsection{Non-uniform intensity of the dynamical diffusion swarm}

The coefficient at the Laplace operator in the diffusion equation equals the intensity of the diffusion. The intensity determines the number of samples passing through the unit square in the unit time. For the simulation of quantum dynamics, we need the diffusion process with the non-uniform intensity. It means that the intensity must depend on the chosen grain $\d x$ of spatial resolution. 

At first, we take up the case when the potential is constant: $grad\ V=0$.
The diffusion process with the non-uniform intensity can rest on the peculiar imaginary mechanism, which we call threads. We illustrate it on the following example. Let us suppose that all samples move not in all space but along some closed curve (thread) determined by the isomorphic injection of the circle to the space: $\g :\ S^1\ar R^3$. We assume that the process of exchange of speeds goes along this trajectory and samples always stay on it, with only change in their linear density. It is equivalent to the imposition of the holonomic bound to samples. We suppose that the linear density and the speed of samples are almost constant in each point of this trajectory. We consider one cube with the points from the trajectory. The stream of samples through its border does not depend on its side $\d x$, because there is only one trajectory. This form of the diffusion process has thus the intensity proportional $1/(\d x)^3$, because the quantity of samples penetrating in the unit of time into the cube with the side $\d x$, is independent of $\d x$, and the density is the ration of the quantity of samples to the volume. This example is not completely appropriate because many areas turns to be without samples at all, but it can serve as the approximate model when the total number of samples is small.

\subsection{Equivalence of quantum and diffusion swarms}

Here our aim is to prove that the sequence of diffusion swarms is the admissible approximation of quantum evolution. We have defined a quantum swarm as the satisfying the equation \ref{swarm}, which evolution is reducible to local movement of samples.  

At first we prove that there exists a quantum swarm, e.g., that it is possible to satisfy \ref{swarm} shifting samples locally. Then we prove that the mechanism of the movement of samples coincides with the diffusion, which gives the main result. 

We consider the quantum swarm, starting from Shredinger equation
\begin{equation}
ih\frac{\partial\Psi(r,t)}{\partial t}=-\frac{h^2}{2M}\Delta\Psi(r,t)+V_{pot}(r,t)\Psi(r,t),
\label{Sh}
\end{equation}
which we can rewrite as 
\begin{equation}
\begin{array}{lll}
&\Psi^r_t(r)&=-\frac{h}{2M}\Delta\Psi^i_t(r)+\frac{V_{pot}}{h}\Psi^i(r),\\
&\Psi^i_t(r)&=\frac{h}{2M}\Delta\Psi^r_t(r)-\frac{V_{pot}}{h}\Psi^r(r)
\end{array}
\label{Sh2}
\end{equation}
for the real and imaginary parts $\Psi^r,\ \Psi^i$ of the wave fnction $\Psi$. We take an interest only in the evolution of the density of quantum swarm, e.g. the function 
$$
\rho(r,t)=(\Psi^r(r,t))^2+(\Psi^i(r,t))^2.
$$

We fix the value $\d x$ and apply for the approximation of the second derivatives the difference scheme of the form
$$
\frac{\partial^2 \Psi(x)}{\partial x^2}\approx\frac{\Psi(x+\d x)+\Psi(x-\d x)-2\Psi(x)  }{(\d x)^2}
$$
for each tine instant, treating the wave function as satisfying all the conditions for the accuracy of this approximation. Since the addition of any constant to the potential energy $V_{pot}$ does not influence to the quantum swarm evolution, we can regard instead of $V_{pot}$ the other potential of the form $V=V_{pot}+\a$, where $\a=-\frac{3h^2}{m(\d x)^2}$, which eliminates the summand $2\Psi(x)$ in the difference schemes for the second derivatives on $x,y,z$ (and thus gives the coefficient $3$) after its substitution in Shredinger equation. To simplify the notations we introduce the coefficient 
$$
\g = \frac{h}{2M}\frac{1}{(\d x)^2}.
$$
Since we yet do not know the mechanism of movement of the samples in quantum swarm, we suppose that we either take some quantity of samples from each cube, or place them in cubes from some storage. We divide the quantum swarm evolution to so small segments of the duration $\d t$, that on each of them samples shift in the framework of neighbor cells only. If we prove that the diffusion mechanism gives the quantum swarm evolution on such a segment, it will be true for the whole evolution because our supposition about the exchange between only neighbor cells does not limit the generality. We also assume that cells differ one from the others only in shifts on $\d x$ along the axes $x$, which also does not limit the generality. We denote the centers of these cells by $x$ and $x_1=x+\d x$. Then due to our supposition about the exchange the summand $\Psi(x-\d x)$ in the difference scheme also disappears and on the considered small time period the following system of equations determines the quantum swarm evolution:
\begin{equation}
\begin{array}{lll}
&\Psi^r_t(x)&=-\g\Psi^i(x_1)+V(x)\Psi^i(x),\\
&\Psi^i_t(x)&=\g\Psi^r(x_1)-V(x)\Psi^r(x),
\end{array}
\label{quant}
\end{equation}
and the analogous system obtained by the replacement of $x$ by $x_1$ and vice versa. This system is applicable in the supposition that samples move from $x$ to $x_1$. If they move from $x_1$ to $x$ the analogous system arises, with the replacement of $x$ by $x_1$ and vice versa. Without loss of generality we can choose the periods $dt$ so small that on each of them movement goes only in one of these directions. In view of the complete symmetry, it would then suffice to consider the case of movement from $x$ to $x_1$; the second case is analogous. Shredinger equation then expresses the evolution process consisting of these two cases, where $x$ and $x_1$ can dispose by six different ways along three coordinate axes.
Now by the period $dt$ we mean the small period when the movement goes only from $x$ to $x_1$.

For this period we have 
\begin{equation}
\begin{array}{lll}
\frac{\partial\rho(x)}{\partial t}&=2\Psi^i(x)(\g\Psi^r(x_1)-V(x)\Psi^r(x))&+2\Psi^r(x)(-\g\Psi^i(x_1)-V(x)\Psi^i(x))=\\
&=2\g (\Psi^i(x)\Psi^r(x_1)-\Psi^r(x)\Psi^i(x_1))&=-\frac{\partial\rho(x_1)}{\partial t}.
\end{array}
\label{1der}
\end{equation}
It means that the loss of samples in one cell equals their return to the other that is the quantum swarm evolution satisfies the condition of locality. Now to compare its evolution with the diffusion we find the second derivative:
\begin{equation}
\begin{array}{ll}
&\frac{\partial^2\rho(x)}{\partial t^2}=2\g [ (\g\Psi^r(x_1)-V(x)\Psi^r(x))\Psi^r(x_1)+\Psi^i(x)(-\g\Psi^i(x)+V(x_1)\Psi^i(x_1))-\\
&(-\g\Psi^i(x_1)+V(x)\Psi^i(x))\Psi^i(x_1)-\Psi^r(x)(\g\Psi^r(x)-V(x_1)\Psi^r(x_1))]=\\
&2\g^2(\Psi^r(x_1))^2-2\g V(x)\Psi^r(x)\Psi^r(x_1)-2\g^2(\Psi^i(x))^2+\\
&2\g V(x_1)\Psi^i(x)\Psi^i(x_1)+2\g^2(\Psi^i(x_1))^2-2\g V(x)\Psi^i(x)\Psi^i(x_1)-\\
&2\g^2(\Psi^r(x))^2+2\g V(x_1)\Psi^r(x)\Psi^r(x_1)= \\
&2\g^2((\Psi^r(x_1))^2+(\Psi^i(x_1))^2-((\Psi^r(x))^2+(\Psi^i(x))^2))+\\
&2\g [(V(x_1)-V(x))((\Psi^r(x))^2+(\Psi^i(x))^2)+o(\d x)],
\end{array}
\end{equation}
где $o(\d x) = (\Psi^r(x)\Psi^r(x_1)+\Psi^i(x)\Psi^i(x_1)-((\Psi^r(x))^2+(\Psi^i(x))^2))(V(x_1)-V(x))$. We now compare it with the expression for the second derivative of the diffusion swarm density found earlier, taking into account that in our case the exchange goes only between the neighbor cells along the axes $x$. Comparing with \ref{swar} and using \ref{intens}, we conclude that the second derivative of the quantum swarm density approximates the second derivative of the diffusion swarm density. 

If we choose as the initial state the state with the density of Gauss form, which coincides with the density of the ground state for oscillator, with the corresponding energy $V=a(x^2+y^2+z^2)$ we will have $\partial\rho /\partial t=0$ in the starting moment in all points. By the proved, the second derivative of the diffusion swarm and quantum swarm densities approximately equal since the diffusion swarm serves as the good approximation of the quantum density on some interval $\Delta T$. Including some potential slowly, we obtain the approximation for all quantum evolution in the limit of swarms when $n$ goes to infinity. 

The swarm approximating one particle quantum dynamics depends on the choice of $\d x$. After this choice we obtain the approximation of the wave function with the error of the order $\d x$ provided $\d t$ we choose as small as required. The intensity of the diffusion depends on the chosen grain, namely it is $\frac{h^3}{m^3c(\d x)^3}$. If we then want to decrease the step of spatial resolution, we must admit the more portion of moving samples in the unit of the volume. It means the uncertainty relation “coordinate – impulse”: the dispersion of impulses of samples grows when $\d x$ decreases. In any case, for the obtaining of the dynamical picture we must fix the grain of spatial resolution $\d x$. 

If the total number $n$ of samples is limited we obtain the model of quantum dynamics with decoherence of the inbuilt type. Such a model we can generalize to the many particle case, so it serves the approximation of quantum dynamics in the standard Hilbert formalism. (see (\cite{Oz1}). The appropriateness of this scheme for the numerical models follows from that it gives the direct probability space for Born rule, in contrast to Copenhagen approach, which takes Born rule as an axiom. 

\subsection{Restoration of wave function from the dynamical diffusion swarm}

We have solved the problem of approximation of the dynamics of quantum particle density by the special diffusion process with the non-uniform intensity. A pair \ref{pair} determines a state of the dynamical diffusion swarm. Such a pair does not use the notion of complex numbers, which generate quantum interference in the standard formalism, and it does not give the beautiful differential equations of Shredinger type for 
  $\rho$ and $\bar p$. Moreover, the mechanism of the dynamical diffusion we introduced for the simulation of quantum evolution differs from the classical processes (as oscillations or heat transfer) in that its intensity depends on the chosen grain of the spatial resolution. We put up with it for the sake of the main: the economy of the computational resources required for the simulation of quantum dynamics. 

Now for the completion of the picture we have to solve the converse task: to chow how to restore the conventional wave function $\Psi$ given a state of the diffusion swarm \ref{pair}. 
For this we address to the equality \ref{1der}, and substitute in it the expression fro the wave function through the density: $\Psi (r)=\sqrt{\rho (r)}\exp(i\phi (r))$. We need to find the phase $\phi (r)$ of the wave function. We note that since only relative phase between points has the physical sense, we can fix some point $r$ and consider a phase in some other point $r_1$ relatively to $r$. If $r_1$ close to $r$, the equation \ref{1der} gives 
$$
\phi(r)-\phi(r_1)=arcsin\ k(\d x)^2\frac{\bar p(\bar r-\bar r_1)}{\sqrt{\rho(r)\rho(r_1)}}
$$
that leads to the following formula for the determining of the relative phase:
\begin{equation}
\phi(r_1)=\int\limits_\g k(\d x)^2\bar v\ d\bar\g
\label{phase}
\end{equation}
where the path $\g$ goes from $r$ to $r_1$. This definition explicitly depends on the choice of the path $\g$, hence we have to prove its correctness, e.g. the independence of this choice. Since the phase is determined within an integer multiplier of $2\pi$, the different choices of the path can result at most in the change of the phase to this number, as for the excited electron state in a hydrogen atom with the nonzero momentum (for example, $3d$). We show that the integration of the speed $\bar v$ of the swarm along the closed contour preserves its value in the time the more exact the less is the grain of spatial resolution $\d x$. If then the definition (\ref{phase}) was correct in the initial time instant, it preserves its correctness for the further evolution as well. 

We thus consider the derivative of the integral of the speed of swarm along the closed contour $\g_c$. Applying the formula (\ref{swar}) in view of $\partial\bar p /\partial t=\rho\ \partial\bar v/\partial t$, we obtain
\begin{equation}
\frac{\partial}{\partial t}\int\limits_{\g_c} (\d x)^2\bar v\ d\g=-\int\limits_{\g_c}I(\d x)^2\frac{grad\ \rho}{\rho}+\kappa (\d x)^2\ grad\ V.
\end{equation}
The first summand gives zero after the integration along the closed contour because it is $grad\ \ln\rho$, the second summand gives zero by the analogous reason.  

It would now suffice to check the correctness of the definition (\ref{phase}) in the starting moment, which we can do for each concrete problem. In the case when the wave function of the initial state comes from the ground state in Coulomb field where $\bar v=0$, the correctness follows form the proved because here there is no phase shift to $2\pi k$. If to find the initial state in the considered problem we need to start from some excited state with some storage of the phase, we need to check the correctness of the definition for this state first. 

\subsection{The case of many quantum particles}

We consider swarm representations of $n$ quantum particles $1,2,\ldots,n$, where $S_1,S_2,\ldots,S_n$ are swarms of samples corresponding to their states $|\Psi_1\rangle, |\Psi_2\rangle, \ldots, |\Psi_n\rangle$. If we consider the ensemble consisting of all these samples, it represents the non-entangled state of the form $|\Psi_1\rangle\bigotimes |\Psi_2\rangle\bigotimes \ldots\bigotimes |\Psi_n\rangle$. However, to represent entangled states of the form 
\begin{equation}
\Phi\rangle = \sum\limits_{j_1,j_2,\ldots,j_n} \la_{j_1,j_2,\ldots,j_n}|j_1,j_2,\ldots,j_n\rangle
\label{many}
\end{equation}
We need to introduce the new essential element in the method of collective behavior. It is bonds between samples from the different swarms. Any basic state $j_i$ we can treat as the coordinates of some particle $i$ in the corresponding configuration space. The representation of the wave function in the form \ref{many} means the existence of bonds connecting points $j_1, j_2,\ldots,j_n$ in the same cortege. The relative quantity of such type of bonds (their number divided to the total number of bonds) is $|\la_{j_1,j_2,\ldots,j_n }|^2$. 

We assume that bonds connect not spatial points but samples of the different real particles. These bonds we write as corteges 
\begin{equation}
\bar s =(s_1,s_2,\ldots,s_n)
\label{cortege}
\end{equation}

where for each $j=1,2,\ldots,n\ s_j\in S_j$. The wave function $|\Phi\rangle$ then acquires the form of the set $\bar S$ of corteges $\bar s$ so that for all $j=1,2,\ldots,\ s_j\in S_j$ there exists exactly one cortege of the form \ref{cortege}. Each cortege plays the role of the so-called world in Everett many world interpretation of quantum theory.\footnote{Strictly speaking, for this we must treat the times $t$ as the part of corteges (see \cite{Oz}). In the non-relativistic theory there is the single time $t$.} We treat this cortege \ref{cortege} as the probe representation of the system of $n$ particles and assume that all interactions go inside one cortege only, whereas the state of the real system resulted from the interference of amplitudes of all corteges occurring in this spatial cell, or resulted from the other process replacing the interference of amplitudes. In the case of one quantum particle, we saw that the exchange of impulses in the diffusion dynamical swarm could serve as such a process. We now study the generalization of this process to the case of $n$ particles. We call $\bar S$ the swarm representation of the system of $n$ particles.

The density of swarm $\bar S$ is given by the formula 
\begin{equation}
\rho_{\bar S}( r_1,r_2,\dots,r_n)=\lim\limits_{dx\ar\infty}\frac{N_{ r_1,r_2,\dots,r_n,\ dx}}{(dx)^{3n}},
\label{density}
\end{equation}
where $ N_{ r_1,r_2,\dots,r_n,\ dx}$ is the total number of corteges locating in $3n$ dimensional cube with the side $dx$ and the center $ r_1,r_2,\dots,r_n$. 

If the wave function $|\Phi\rangle$ is the tensor product of one particle wave functions:
$$
\Phi\rangle = \bigotimes\limits_{i=1}^{n}|\phi_i\rangle 
$$
then the corresponding bonds we can obtain by the random choice of samples from the uniform probability distribution $s_j\in S_j$ for each $j=1,2,\ldots,n$, which thus form each cortege $s_1,s_2,\ldots,s_n$.
With such a choice of corteges we obtain that the density of the corresponding swarm satisfies Born rule, which we can read for swarms in the form 
\begin{equation}
\sum\limits_{\bar r\in D}|\langle\bar r|\Phi\rangle |^2 = \frac{N_{\bar r, \bar S}}{N}
\label{born}
\end{equation}
where $D\subset R^{3n}$, $ N_{\bar r, \bar S}$ is the total number of corteges located in the area $D$. For the entangled state $|\Phi\rangle$ this choice of corteges for the set of swarms $\bar S$ does not give us the condition \ref{born}. We thus have to take \ref{born} as the definition of the corteges in $\bar S$. To determine the swarm we must also define the speeds of all samples, namely, to generalize our construction to the case of $n$ real quantum particles.   

Let $\Psi(r_1,r_2,\ldots,r_n)$ be the wave function of the system of $n$ particles, 

$\Psi=|\Psi|exp(i\phi(r_1,r_2,\ldots,r_n))$ is its Euler representation. We denote by $\grad_j\phi(r_1,r_2,\ldots,r_n)$ the gradient of $\Psi$, taken on coordinates of the particle $j$, where $j\in\{ 1,2,\ldots,n\}$ is the fixed number. For $n$ particles the formulas of the relation between the wave and swarm representation of the quantum state have the form 
\begin{equation}
\begin{array}{ll}
&|\Psi(\bar r)|=\sqrt{\rho(\bar r)};\\
&\phi(r)=\int\limits_{\bar \g :\ \bar r_0\ar \bar r}k(dx)^2\bar v\cdot d\g,\\
&\bar v = a(dx)^{-2}\bar \grad\ \phi(\bar r),
\end{array}
\label{connection_n}
\end{equation}
where $\bar r$ denotes $r_1,r_2,\ldots,r_n$, $\bar\grad$ denotes $(\grad_1,\grad_2,\ldots,\grad_n)$, and $\bar \g$ is the path in $3n$ dimensional space. The rules \ref{connection_n} are sufficient for the determining of the swarm given a wave function, provided we agree to join samples into corteges independently of their speeds. The microscopic mechanism of the swarm dynamics then acquires the following form. The exchange of impulses between two corteges of samples: $\bar s=(s_1,s_2,\ldots,s_j,\ldots,s_n)$ and $\bar s’=(s’_1,s’_2,\ldots,s’_j,\ldots,s’_n)$ is the impulse exchange between two samples $s_j$ and $s’_j$ provided $\bar s$ and $\bar s’$ belong to the same spatial cube in the configuration space $R^{3n}$ for $n$ particles. Here $j$ is chosen randomly from the uniform distribution. With this definition the reasoning we earlier carried out for one particle can be repeated literally, and we obtain that this microscopic mechanism of impulse exchange for $n$ particles ensures the approximation of $n$ particle quantum dynamics within the accuracy of the order $dx^{3n}$ for the definition of the wave function on the fixed time period. 

The method of collective behavior gives us the basement for the economical modeling of the real quantum evolution, including the unitary evolution and decoherence. 

\newpage

\section{Lecture. Heuristic of collective behavior}

We have established that the simulation of quantum evolution by swarms with samples joint in non-overlapping corteges gives the approximation of the unitary quantum evolution provided the quantity of samples of each particle goes to infinity. Factually, the model of a quantum system with many particles we represent in the form of two-dimensional net of the form shown at the picture. Here in each column stand samples of one quantum particle and in each row stands the quantum world, to which from each real particle belongs exactly one sample. Quantum exchange of impulses goes on each real particle between close worlds, and the potential energy acts in the framework of one world only, e.g., each sample is subject of influence from only those samples, which belong to the same world with it. 
\begin{figure}
\centering
\caption{Net representing the system of $m$ quantum paritcles}
\vspace{150mm}
\makebox[380mm][l]{\includegraphics{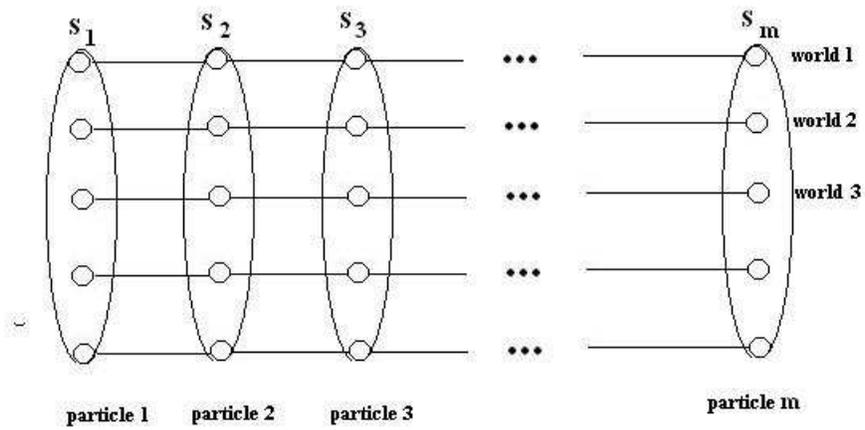}}%
\end{figure}

The dynamics determined by this net differs from the unitary only to the extent of the total number of corteges (equal to the quantity of samples of each particle) is not infinite. If the quantity of particles grows the rapid divergence of the net dynamics from the unitary, determined by Shredinger equation takes place.  It is just the absolute model of decoherence, when the lack of classical memory for the keeping of quantum states plays the role of decoherence. However, it is not sufficient for the building of modeling algorithms. We must somehow determine how many samples of real particles do we need, e.g., what is the real value of $n$. We can treat the determining of the real value of $n$ in the nature represents the purely physical problem admitting the exact solution. I do not know it. Though we can assert that the value of $n$ can come only from the observing of the dynamical scenarios built by the net. Moreover, this value can depend on the simulated process in the following sense. Processes with the small length of corteges $s_1,s_2,\ldots, s_m$ can allow the bigger values of $n$, whereas the process with the large $m$ conversely presumes the small values of $n$. In any case, the capability of the dynamical scenario influences to it much more than the known physical constants. 

What new the swarm representation gives for the description of many particles comparatively to the standard quantum formalism? The new possibility is that the entanglement obtains the direct representation in the form of corteges $s_1,s_2,\ldots, s_m$, which consist of separate samples of all particles connected by the bonds. We call the bonds the objects, which connect the samples to the corteges – quantum worlds. Bonds have no analogs in the standard formalism, which also has no samples. The standard treatment defines the entanglement as the impossibility to represent the state vector as the tensor product of one-particle states. It is no way to work with this definition “from the contrary” but the matrix algebra, where decoherence is the foreign element, which thus does not admit any analysis from the first principles. Nevertheless, it is reasonable to treat the entanglement as the physical resource, as particles themselves. It means that we must ascribe to the entanglement some formal object making possible to manipulate with it directly. Only constructivism gives this possibility. In the heuristic of collective behavior, we can introduce rules of the break of a long bond when the (conditional) burden to this bond exceeds some threshold. The other form of the work with bonds is their selection. The selection of quantum states has the following form. Corteges – worlds we can join to the separate groups $\G_1, \G_2,\ldots, \G_l$ so that any group will contain corteges close to each other in the configuration space for $m$ particles. We then can cancel groups with small quantity of samples and redistribute bonds between the rest samples accordingly to the closeness to worlds in large groups. This principle of quantum selection has the interpretation in terms of constructiveness of the amplitude interference. 

We consider $d$ samples each of which carries the amplitude of the module $\a$, but with the different phases. If these samples stay near each other, their phases are close and the deposits of these samples in the resulting probability are approximately $(d\a )^2$, because we must add just their amplitudes. If these samples are far, their deposit to the probability is $d(\a )^2$, which is in $d$ times lesser, because now we have to add probabilities (phase are distributed randomly).  

This reasoning is valid for the electromagnetic field as well. Given an ensemble of atoms, which can emit photons, we have two possibilities:

\begin{itemize}
\item The change of state of the atom, emitting the photon makes possible to identify it among the others atoms of this group. 
\item The change of state of the atom, emitting the photon redistributes among all atoms of the group that makes impossible to identify this atom.
\end{itemize}

Let $\psi$ be the amplitude, which the emitted photon carries, $\Psi$ be the state of the group of atoms. We ascribe the emitting atom the index $j$ and assume that amplitudes $\la_j$ of the emission of each atom $j$ are approximately equal. Let the zero value $j=0$ corresponds to the case when the photon is not emitted. In the first case, the state of the whole system of atoms and photons is determined by the common wave function of the form 
\begin{equation}
\sum\limits_j \la_j|\psi_j\rangle |\Psi_j\rangle ,
\end{equation}
in the second - 
\begin{equation}
\sum\limits_j \la_j|\psi_j\rangle |\Psi\rangle = |\psi_{general}\rangle |\Psi ,
\end{equation}
where $|\psi_{general}\rangle =\sum\limits_j\la_j|\psi_j\rangle$ is some photon state. We estimate the probability of the photon emission in these cases. In the first it is 
\begin{equation}
\sum\limits_{j\neq 0}|\la_j|^2,
\end{equation}
in the second
\begin{equation}
|\sum\limits_{j\neq 0}\la_j|^2.
\end{equation}
Since all $\la_j$ for $j\neq 0$ are approximately equal, the probability of the photon emission in the first case is much lesser than in the second: the coefficient is the total number of atoms. This qualitative reasoning belongs to Feynman (see \cite{Fe}). Its construction interpretation rests on the notion of bonds between samples of atoms and photons. In terms of collective behavior, the redistribution of the return impulse between atoms in the photon emission goes through the bonds between samples of atoms. In standard terms, atoms must be in the entangled state; however, the detailed description of the redistribution is hardly possible, because just such problem requires a quantum computer. The constructive heuristic allows the simple reasoning. The entanglement means the presence of stable bonds between the samples of separate atoms. The redistribution comes from the kind of elastic forces generated by bonds in this world. This property of elasticity is not the additional element of the reality, but the main property opf bonds; for example, it ensures the interference properties of a molecule which wave function has GHZ type: $\sum\limits_{\bar j} \la_{\bar j}|\Psi_{j_1}\rangle|\Psi_{j_2}\rangle\ldots |\Psi_{j_k}\rangle$, where $|\Psi_j\rangle$ are states of its atoms. 

The existence of bonds between atoms thus substantially increases the probability of the photon emission. We study the difference between the separate photons and the electromagnetic field. This difference is that the field is the ensemble of photon samples connected by bonds. In the standard formalism the photon wave function – in our example $|\psi_{general}\rangle$ represents (after the appropriate renormalization) the intensity of the classical electromagnetic field. In the first case, when we have the separate photons, we can speak about the field only conditionally, e.g., after the detecting of the concrete photon and this filed if very small. In the second case, when the return impulse from the photon is redistributed, we have the solid value – photon amplitude, which reveals classically. This is the standard description of the photon radiation. We see its duality: in the case of the field we cannot speak about the probability, but only about the field, measurable classically, whereas in the case of separate photons we speak just abut the probabilities. The standard approach makes impossible to use the same language for the description of the emission in the cases of separate photon and the field. This is the feature of standard formalism. 

In the constructive formalism the both situations – with the separate photons and the field have the uniform description. The field differs from separate photons only in the presence of bonds between samples. What we earlier called the elastic properties of these bonds guarantees the possibility to observe the field by classical means. It is important to stress that bonds are not material objects; it is only the mathematical abstraction belonging to the constructive version of many particle quantum mechanics. It is thus senseless to ask questions about their mass, etc. Analogously, what we call the elasticity of bonds is not the usual elasticity obeying Guk law. It is the heuristic trick only, which helps to solve the main problem of constructive quantum theory: the creating of the working model of dynamics. 

This situation shows why the detection of the entanglement of quantum states of a complex system can be difficult. The entanglement is not the characteristic firmly connected with the fixed particles, but can pass to the others particles.  Let us consider the system of three qubits in the standard formalism, when they are in the states $\frac{1}{\sqrt 2}(|0\rangle+|1\rangle ),\ |0\rangle ,\ |0\rangle$ correspondingly. The sequential operations $CNOT$ at first on the first two particle, then on the first and third give the state of the form $\frac{1}{\sqrt 2}(|000\rangle+|111\rangle )$, in which all three qubits are entangled. If we now apply the next operator $CNOT$ to the first and second qubits it results in the entanglement of the first and the third qubits whereas the second will be non-entangled with them. The entanglement can thus easily pass to the other particles. In the various interactions in the complex system, this situation is the rule and it leads to the difficulty in the detection of the entanglement. In any case, in the complex systems the methods of detection of the entanglement working for ions in the trap or for biphotons do not work. 

Nevertheless, the general scheme of the detecting of entanglement is also applicable to complex systems, provided we have many identical copies of their states. This scheme is the quantum tomography. Given a state of two qubits $S_1$ and $S_2$, we must determine is it entangled or not. For this we need to prepare this state twice and measure it in the basis $E=E_1\bigotimes E_2$ and in the basis $E'=E'_1\bigotimes E'_2$. If $E_1$ coincides with $E_2$ and $E'_1$ coincides with $E'_2$, the results of measurements will be correlated. For example, we can thus distinguish EPR state $\frac{1}{\sqrt 2}(|00\rangle+|11\rangle )$ from any mixed state with non-entangled components. We can vary this scheme and fit it to more complicated systems $S_1$, $S_2$, we also can choose two basic systems $E_1$ and $E'_1$ differently. The main is that in the case of the same choice of these basic vector systems the results of measurements must be correlated. In the constructive quantum mechanics, this property is the characteristic of bonds between samples of the different particles: we must take into account the reaction of the whole quantum world in the interactions between samples. Of course, it does not mean that the reaction of samples connected by bonds must be equal; it may be opposite. The main is the presence of some correlation in the reactions of all samples, belonging to this quantum world. 

The various examples of such correlations are well known. We consider two living things and their reaction to the same external influence. If this influences are graduated so that we can compare the reactions of the things to these influence, these reactions will be the more similar, the closer these things are.  We can introduce the degree of closeness of living things by their DNA closeness. The degree of closeness of their reactions will be then in the direct dependence of the degree of their affinity. This closeness may be almost absolute, as for the one-ovum twins. It brings the analogy with the entangled quantum states in which different objects may be. However, the quantum entanglement is not reducible to classical correlations. It follows from the violation of the so-called Bell inequalities, which we discuss later. The peculiar sense of quantum correlation is that the nature behaves as the constructive model, in which the user cannot interfere into the created video film, but must observe it from the beginning to the end. Its free will permits only to order to the system to create such a video film. In the process of building of it, the communications between spased points takes place, which the user can treat as the instantaneous communications. It is necessary to show the quantum correlations in the film. These communications belong to the administrative part of the model, inaccessible for the user. In particular, the user cannot apply these communications for the transmitting the information he created to the far distance, so that the principle of relativism remains valid.  If we assume this limited treatment of the free will as the natural, no obstacles for the constructive representation of the entanglement in the form of bonds remain. This representation is the direct analogy of the quantum entanglement with the classical connections; we must accept this treatment if we want to enter the world of complex systems. 

In the standard formalism nothing stays behind this analogy and it is the superficial similarity. Indeed, in any reasonable specification of the term “hidden parameters” it is possible to prove their absence in quantum theory (one version of it is Cohen Spector theorem, there are the others). It means the following. In the standard formalism there is nothing but the wave function that affects the result of the measurement. This postulate of quantum mechanics is called the theorem about the absence of the hidden parameters. 

The theorem about the absence of hidden parameters has the sense only in the standard formalism. Namely, the sense of this theorem is that the space of elementary events for quantum probability is inaccessible for us. It the constructive quantum theory we always can use explicitly elementary events operating with corteges of samples, which compose quantum worlds. It gives us not only the direct form of elementary events for quantum probability but also the possibility of the immediate manipulation with non-local states in complex systems, which role is yet to be elucidated. 

\subsection{About fermionic anti symmetry of wave function}

We consider the representation of the fermionic symmetry of the wave function in terms of collective behavior. Let $|\psi_j\rangle$ denote one particle wave functions, and $|i\rangle$ denote its spatial positions. Accordingly with the requirement of the anti symmetry of the wave function of the system of the independent identical fermions it must have the form $|\Psi\rangle=det\ (\langle\psi_j|i\rangle)$, where in the expansion of the determinant all multiplications we treat as tensor products. Such states $|\Psi\rangle$ we take as the non entangled states of the system of identical fermions, whereas all other states can be expanded to the rows on these functions taken for the different choices of one particle $|\psi_j\rangle$. From the view point of Hilbert formalism, such states are entangled. 
   However, in the representation of collective behavior we can easily make them non-entangled. Really, imagine that the supports (area where the function is non-zero) of them do not overlap. Then the determinant becomes the product of one-particle functions. We can make the supports non-overlapping by the decreasing of the grain of spatial resolution $dx$. In the swarm representation, two swarms cannot coincide even in one point due to the Pauli principle, and the determinant we can then replace by the ensemble of two swarms. By the way, it makes possible to specify the particles, which are identical from the canonic viewpoint. This specification follows from the completely different supports of their wave functions. I stress that in the collective behavior the fermionic character of particles belonging to the same type follows from Pauli principle and has the representation as the non-overlapping supports of their wave functions (not intersecting swarms).

\subsection{Constructive viewpoint to quantum electrodynamics}

We look how to make constructive the formalism of quantum electrodynamics. The ordinary (non-relativistic) quantum theory deals with the wave function dynamics $|\Psi (t)\rangle$, where the physical time $t$ is the common parameter; the case of relativism is the different. Here we have a state $|\bar\Psi\rangle$, which is in the following relations with the time. There is the so-called internal time of quantum system $t$, and there is the external time $\tau$. The evolution accessible for the user observation is the sequence of the unitary operators in the space of states ${\cal H}$ of the form
\begin{equation}
|\bar \Psi_{\tau=0}\rangle\ar|\bar \Psi_{\tau=1}\rangle\ar\ldots\ar |\bar \Psi_{\tau=l}\rangle
\end{equation}
called scatterings. This is the chain of the quantum state evolution in the external time $\tau$, which periods we denote by integers. We choose them so that this sequence gives us the realistic picture, for example, of the chemical reactions. Here all states $|\bar \Psi_{\tau=j}\rangle=|\bar\Psi\rangle$ have the form
\begin{equation}
|\bar\Psi\rangle = \sum\limits_i\la_i|\bar x_i\rangle |\bar t_i\rangle
\end{equation}
e.g., is the linear superposition of basic states $|\bar x_i\rangle |t_i\rangle$. Each of such states means that the considered system is in the point $\bar x_i = (x_i^1,x_i^2,\ldots, x_i^n)$ of the configuration space in the moment $\bar t_i=(t_i^1,t_i^2,\ldots, t_i^n)$ of the internal time. Here $x_i^j$ is the coordinate of the sample $j$ of the real quantum particle $i$ in our system, $t_i^j$ is the time instant corresponding to this sample. The internal time of the system in QED is not thus the unit parameter, but something similar to the spatial coordinate: each sample has its own time. The presence in the expansion just the time cortege $\bar t_i$ follows from the conventional rules of the operations with tensor products of states $|\bar \Psi'\rangle$ and $|\bar \Psi''\rangle$ for the different particles, when their state in a non entangled case has the form $|\bar \Psi'\rangle\bigotimes |\bar \Psi'\rangle$, and in an entangled state is the superposition of these states. Non relativistic states, in which the time is common is factorized as the tensor multiplier, can be only spatially entangled, whereas the relativistic states have no common internal time, and states can be entangled on the internal time as well. 

\begin{figure}
\centering
\caption{Network representing scattering}
\vspace{150mm}
\makebox[380mm][l]{\includegraphics{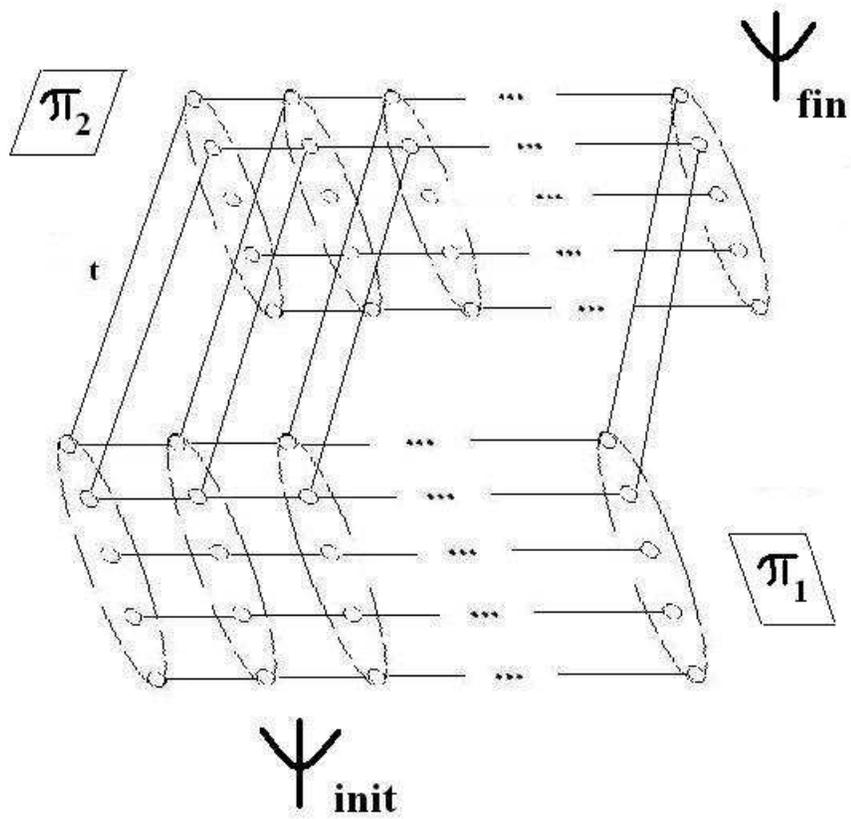}}%
\end{figure}

It is possible to prove (see \cite{Oz}) that the law of energy – impulse conservation in the fundamental interaction between charged particles and photons gives the presence of the common internal time for the interacting particles within the chosen grain. It means that inside each quantum world for close samples the time must be approximately the same because they must interact. Analogously, for the samples of the closed quantum worlds (here the closeness include the times $t_i^j$ as the coordinate in the coordinate in the configuration space) exchanging by impulses the difference in the time must be in the framework of the grain. We obtain the following picture for the relativistic net of QED corresponding to one state $|\bar \Psi\rangle$. Here the representation will be not two dimensional, as in the non-relativistic case, but three dimensional, because the internal time $t$ now belongs to the configuration space and it has the same detailed structure as the space coordinate $x$. 

What means now the selection of quantum states in the evolution in the external time $\tau$? On each passage from $\tau$ to $\tau+1$ we select states of the form $|\bar x_i\rangle |\bar t_i\rangle$. In is not the selection of ordinary wave functions distributed in the space. This is the selection of dynamical scenarios, e.g., wave functions distributed in the space-time. How this selection goes? Its simplest form is the grouping of states by their closeness in the configuration space-time and the selection of groups with the maximal density. It is the simple generalization of the corresponding principle for the non-relativistic case. In this selection, factually bonds of two types are in use.

\begin{itemize}
\item Bonds, joining samples of the different particles, e.g., bonds of the type we already discussed. 
\item Bonds, joining one sample taken in the different instant of its individual time. 
\end{itemize}

The second type of bonds means the presence of individual histories of each sample. How these bonds on the internal time work in the modeling? They have the certain role. We select not only close states, but close dynamical scenarios. Such scenarios may be the attribute of concrete samples of real particles. Their regrouping in the selection in the genetic algorithm means the recombination of these attributes. The both these types of bonds belongs to the planes $\pi_1$ or $\pi_2$ at the picture.

Are there more complex methods of selections than the grouping of scenarios by their closeness? In the other words, is it possible to introduce bonds corresponding to summands in the wave function of the form $|x_1\rangle |x_2\rangle |t_1\rangle |t_2\rangle$ for the different times $t_1\neq t_2$? Such bonds would have the form of slanting lines in the picture, e.g., non-parallel to the planes $\pi_1$ or $\pi_2$. The introduction of such bonds may come from the new methods of the dynamical scenarios selection; hence, this question remains without answer. 

The constructive consideration quantum system with many particles in QED leads us to the work with dynamical scenarios, each with its internal time $t$. It includes the selection of such scenarios. The surviving of certain scenarios replaces the surviving of the certain states in non-relativistic models. In these scenarios participate many samples of real particles, which can be in the scale of external time for a long time $\tau$, which gives the observed phenomena of the microscopic character.  

\subsection{How can we create quantum computer and why do we need it: realization of collective behavior on computational networks}

The method of collective behavior gives the principal way to overcome the problem of quantum computer, e.g., the exponential growth of computational resources needed for the exact modeling of the unitary dynamics in the full Hilbert space of states. Here we treat decoherence as the fundamental factor, represented as the limitation of the classical memory. It makes possible to simulate biochemistry on classical computers. 

However, in the heuristic of collective behavior we meet the difficulty as well. It is the necessity to transform the information along bonds joining samples of the same quantum world. This difficulty may seem empty to the pure mathematicians, but this is the real problem. The scientific logic joins samples not in quantum worlds but in swarms corresponding to real particles. It makes unavoidable the situation when the different groups of programmers work at the different groups of real particles. If we join real particles in some non overlapping ensembles $\G_1,\G_2,\ldots, \G_k$, the evolution model of each of them $\G_j$ will work at the corresponding processor $P_j$. Bonds joining samples inside each quantum world then become the bonds between these processors. 

What kind of information can these bonds transmit? At first, it may the simple synchronization as the type of classical correlation. This kind of information is similar to the open key distribution in quantum cryptography. Let the different processors $P_1$ and $P_2$ have to generate the common code, which will then work at them. This problem arises in the modeling of the specific for the living thing proteins – gamma globulins allowing the identification of own tissues by antibodies. For this kind of problems, we can use the methods of one particle cryptography, including its quantum version. However, if the synchronization is included in the complex process of local computations on both processors, the preferable choice for it will be bi-photonic mechanism. This mechanism causes no delay in the computation. The idle time of processors coming from the passage of the signal from $P_1$ to $P_2$ may be of the order of hundreds microseconds if these processors are in the different towns, and it can valuably brakes the work of the whole net. This argument touches the organization of computations only, not the modeled process themselves; I do not assert that in the real systems the synchronization is carried out by this mechanism. 

\begin{figure}
\centering
\caption{Penetration of transport molecule through pore in cellular membrane}
\vspace{200mm}
\makebox[250mm][l]{\includegraphics{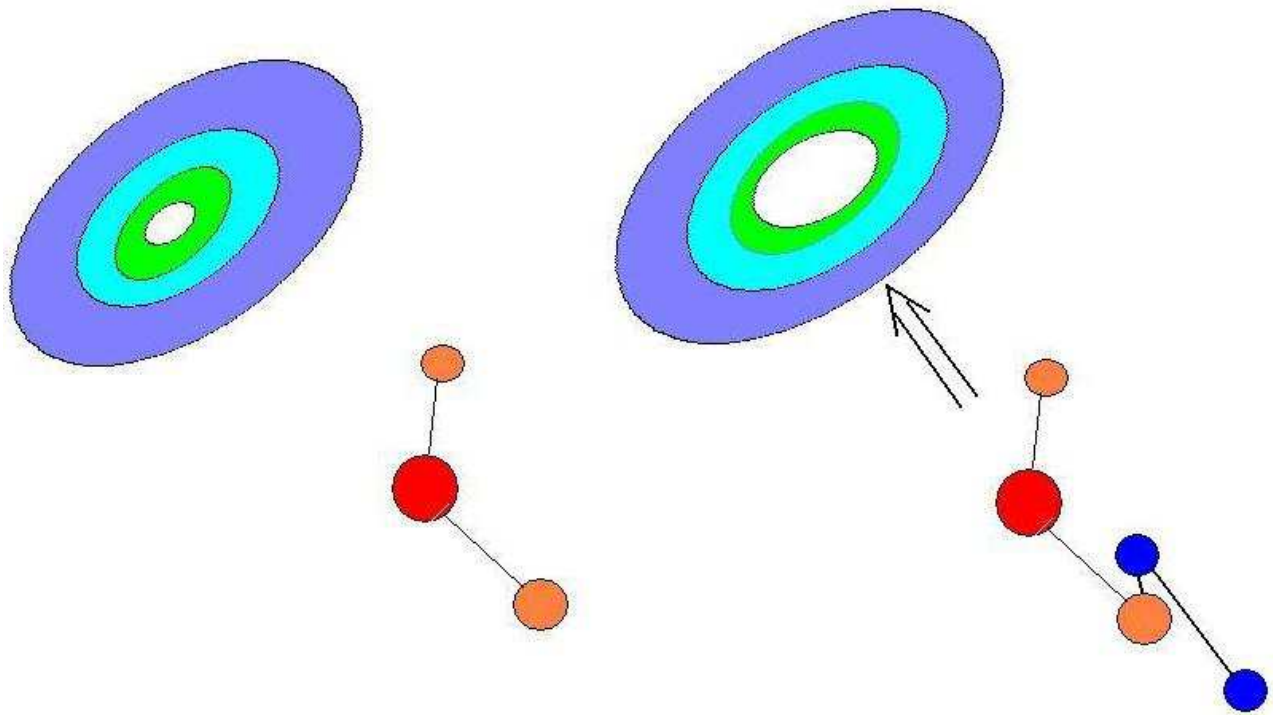}}%
\end{figure}

The other example illustrating the importance of synchronization is the process of penetration of a molecule through the cellular pore. The cellular membrane has pores, which can be either in open or in closed state. Penetrating molecules transport important substances, and at the same time, the membrane must be impenetrable for the other, harmful substances. The pore thus must open just in the moment when this transport is ready for the penetration. The pore itself is the complex mechanism, as the protein molecule, it is thus non-reasonable to charge the same processor by the simulation of the internal evolutions of these two systems. The simulation of the whole process thus requires the synchronization of evolutions for the obtaining of the desired effect. 

I speak here about the instantaneous classical correlation of the different processors by pairs of photons. The usage of entangled photon pairs promises the more advantages. If the forming of hydrogen bonds between two complex molecules belongs to the modeled process, for the simulation of such quantum particle belonging to each system it would be useful to have the real quantum model of this bond in both parts of the computational process. 

Of course, we can simulate each quantum correlation by classical means if admit the possibility of the signal to travel form one processor to the other. The drawback of this ideology is that it contradicts to the basic requirements of the computational net architecture. These requirements set the general form of control shown at the picture. 

\begin{figure}
\centering
\caption{Structure of complex system model}
\vspace{150mm}
\makebox[250mm][l]{\includegraphics{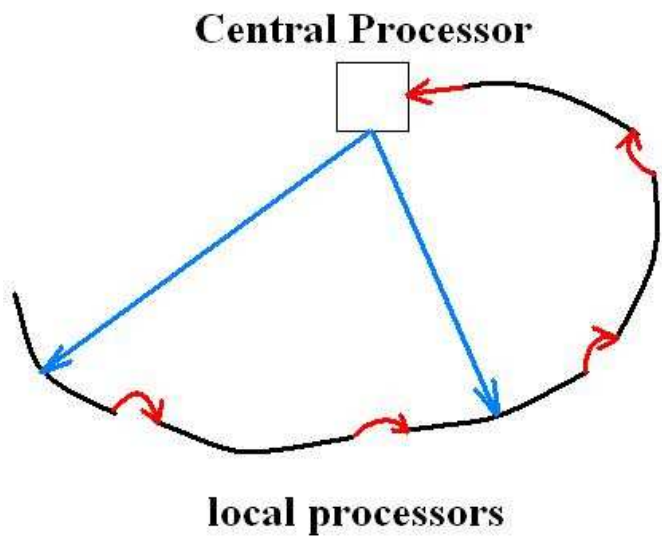}}%
\end{figure}

The central processor sends signals to the different local processors, each of which models the corresponding subsystem in the whole system. For example, CP decides that it is the time for the synthesis of some polymer $A$ of the specific activity in some subsystem and simultaneously – the other polymer $B$, suppressing (or intensifying) this activity in the other subsystem. CP sends the package of signals to the both subsystems, then immediately switches over to the next task (for example, the simultaneous synthesis of the other pair $A'$ and $B'$). What happens if subsystems themselves begin to send signals to each other? Let we be given $m$ different subsystems, ach of which controlled by its own processor. For the right addressing of mutual signals between local processors (about $m^2$ ways to choose them) we need to charge CP with the addressing of these signals. CP then will have to wait the time $cD$, where $D$ is the distance between local processes, $c$ is the speed of light before the switch over to the next task. If there are many tasks (in the real model of biochemical metabolism there are very many such tasks) this delay results in the fatal braking of the model. This is why the idea of the direct exchange of signals between the different local processors is not good. We thus come to the necessity to use one side control from CP, which realizes it by sending of signal packages to the different pairs of processors. CP also receives signals from the local subsystems, but not directly, by the chain of connected processors. This form of the organization of a living system model is more appropriate because we maximally relieve CP from the routine tasks of the metabolism maintenance. 

The architecture of the computational net must then have the form shown at the picture, where the general control realizes CP. It communicates with the bulk of local processors so that the signal form the remote subsystem comes to CP through the chain of intermediate subsystems only. 

The collective behavior heuristic represents the entanglement of quantum states by the special object – bonds between samples of the different real particles. We suppose that just manipulations with bonds makes possible to model complex systems, because this is the only way to work with entangled states. One processor will process bonds between samples belonging to the same subsystem. Bonds between samples from the different subsystems require the special attention. These bonds we call remote. I stress once again that we speak not about the real distant subsystems but about the distant local processes in the net. There are seasons to treat the remote bonds as important for the obtaining of the right scenarios of the complex system dynamics. 
Hence we must charge CP with t he work with the remote bonds. It can act purely classically, and send signals to the chosen local processes.

There exist situations where CP using entangled bi-photons has the advantage over CP with classical signals. 
We consider the following abstract problem. Let we have to synthesize two molecules of polymer
$$
\begin{array}{ll}
&A_1, A_2,\ldots, A_m\\
&B_1, B_2,\ldots, B_m
\end{array}
$$
consisting of mono blocks of the two types: $a$ or $b$ (see the picture). Each mono block has the lateral surface, which can glue if it adjoins to the similar surface of the other mono block, and the internal surface, which can glue to the other mono block only by the special boll in its center (see the picture). This construction ensures the non-symmetric gluing when two mono blocks are shifted either in the same direction, or in the opposite directions. Two glued mono blocks we treat as firmly fastened. Mono blocks are linked by the intermediate linear segments, which can stretch or shrink. Here each mono block can be shifted along the chain to some $dx$ either to right or to left. The synthesis goes by the sequential addition to the existing chain the new mono block, which comes to the point of assembly first (mono blocks in the storage are in the chaotic movements). It is possible to shift a mono block either ahead, or back along the chain to the small distance $dx$. We denote the shift ahead by $+$, the shift back by $-$. Each segment $A_j$ or $B_j$ is the pair of the form $c, s$, where $c\in\{ a,b\}$, $s\in\{ +,-\}$. Let the assembly of the polymer $A$ goes in one point, and the assembly of the polymer $B$ - in the other point so that their models are located far one from the other (for example, they are located in the different countries). The problem is to organize their synthesis so that the quantity of critical impositions in the matching of the chains $A$ and $B$ is as small as possible. We call the critical imposition such imposition, in which the mono blocks are not firmly fastened.

\begin{figure}
\centering
\caption{Imposition of two polymers.
 Red color shows the connections}
\vspace{150mm}
\makebox[380mm][l]{\includegraphics{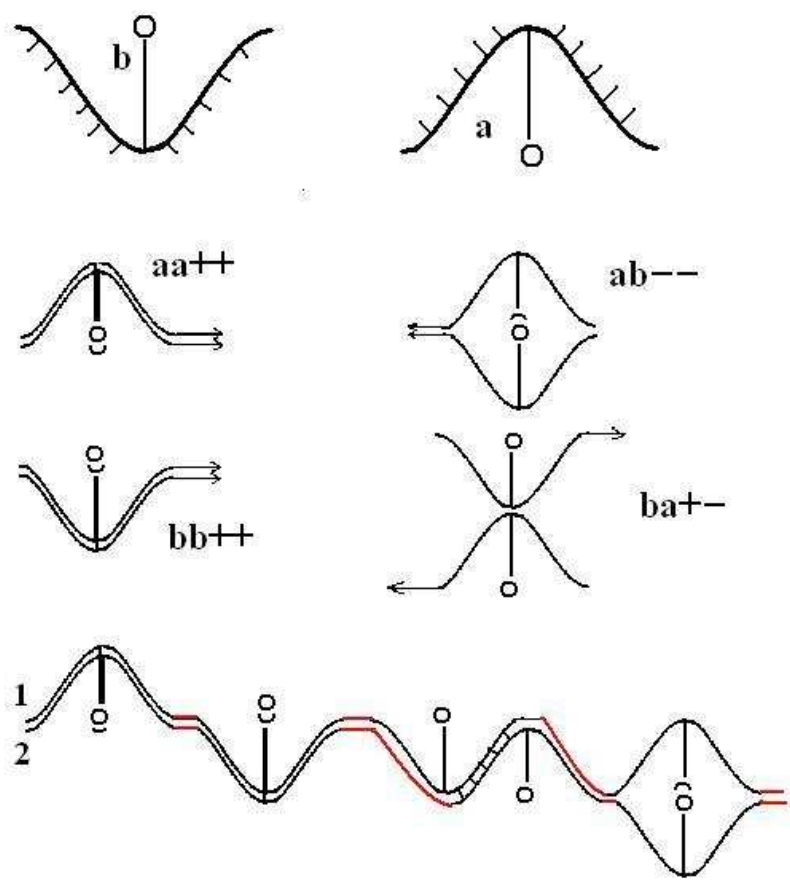}}%
\end{figure}

Such problem could appear in the modeling of the simultaneous synthesis of gen and its anti-gen in the different cells. Of course, we could organize the simulation by the information exchange between the two points of assembly at each step. However, if the corresponding processors are far one form the other this exchange would substantially impedance the modeling because we have to wait when the signal travels between computers at each step of the process that reduces the efficiency of the using of fast computers. It is possible to organize the synchronization by the classical correlated photon pairs. However, the using of the entangled bi-photons in this problem gives the valuable advantage over the classical schemes. 

Since for the minimization of the total number of critical impositions we intend to use the state of EPR type $|\Psi\rangle =\frac{1}{\sqrt 2}(|00\rangle +|11\rangle )$, it is necessary to substantiate that it gives the advantage over the classically correlated photon pairs. For this we recall Bell inequality, valid for all classical correlations. Let we be given four classical random variables $b_2,b_1,a_1,a_2$, taking values from the set $\{ 1,-1\}$, for their mean values Bell inequality takes place: $M(|a_1b_2+b_1b_2+a_1a_2-b_1a_2|)\leq 2$, which follow from the representation  
$a_1b_2+b_1b_2+a_1a_2-b_1a_2=a_1(b_2+a_2)+b_1(b_2-a_2)$ because one of the brackets is zero and the module of the other is 2. 
We accept the agreement. The lower index denotes the number of the point of assembly: 1 or 2. The letter $a$ or $b$ denotes the type of mono block, and the value of the corresponding value equals $+1$ if we shift the mono block ahead, and $-1$, if it shifts to right in the assembling. The result of the assembling in the both points is determined provided for each lower index 1 or 2 we have at first, the letter $a$ or $b$, and at second, the sign of the shift $+$ or $-$. The letter $a$ or $b$ is always determined by the type of the mono block, which at this moment randomly occurs in the closest distance from the assembly point. We thus can control the assembling only choosing signs of the shifts: $+$ or $-$ in the both assembly points. To guarantee that the communication between them does not decelerate the modeling of the assembling process, we need to choose two sings in both points simultaneously. In the case of classical correlation of these choices, we would have Bell inequality. For each step of the assembling process we introduce the index of criticality as $Cr=+1$ if it gives non critical imposition, and $Cr=-1$, in the opposite case. We are interested in the total number of non-critical impositions in the whole polymer $NonCr$. Our aim is to make it maximal. For one pair of mono blocks we have $NonCr=\frac{1}{2}(1+Cr)$. Because each combination $aa, ab, ba, bb$ occurs with the probability $1/4$, for the mean value $M(Cr)$ of the index of criticality we have 
\begin{equation}
M(Cr)=\frac{1}{4}(a_1b_2+b_1b_2+a_1a_2-b_1a_2)
\label{cr}
\end{equation}
which by Bell inequality gives us for the mean value of the number of non-critical impositions the inequality $M(NonCr)\leq \frac{1}{2}(1+\frac{2}{4})=\frac{3}{4}=0.75$. 

In the case of quantum EPR state of bi-photons the situation is the different. There is no the common probability space for two point of assembly, where $a$ and $b$ with indexes are random variables. Of course, there exists the common probability space of elementary events $P$ for quantum probability, but in this space we cannot treat $a$ and $b$ as random variables, because this general space $P$ consists of non-local events. It means that here we have not Bell inequality and must compute the probabilities directly, using Born rule. Let in each assembly point stands the corresponding photo detector, which can obtain orientations corresponding the desired observables $a$ and $b$. For the first and the second points, these observables have the form 
\begin{equation}
\begin{array}{lll}
&a_1=\sigma_x,\ &b_1=\sigma_z,\\
&a_2=\frac{1}{\sqrt 2}(\sigma_z-\sigma_x),\ &b_2=-\frac{1}{\sqrt 2}(\sigma_x+\sigma_z)
\end{array}
\end{equation}
correspondingly. We agree that the type of the current mono block in the assembling determines the corresponding position of the detector in each point, and the shift sing of the mono block equals the value of the corresponding observable. Again, since all combinations of the monomer types $aa, ab, ba, bb$ have equal probabilities we can use the formula \ref{cr} for the mean value of the index of criticality. However, now the mean value of this sum will be $2\sqrt{2}$ (it is found straightforwardly) and for the mean value of the total number of non-critical impositions we obtain the value $M(NonCr)=\frac{1}{2}(1+\frac{2\sqrt{2}}{4})\approx 0.85$. The using of EPR pair of photons thus gives the substantial advantage for this problem. 

The effect we obtain could seem small, but if we repeat the assembly sequentially alternating it with some other processes, the probability can multiply and the effect may become very important. Moreover, this situation despite of its abstract character, in all likelihood is typical, and can appear in the different non-equilibrium processes in a living cell. 

For example, we look at the process of the penetration of a molecule through a cellular membrane. Two abstract factors $a$ and $b$ can determine the usefulness of it, which are the attributes of the cell itself and the penetrating molecule. For example, for the cell, it may be the different states of the molecule-helper, which pulls through the pore the considered molecule; for the penetrating molecule, it may be its chemical status. We associate with the cell the lower index 1, with the molecule – the lower index 2. We assume that the sign determines is the pore open or not; for the molecule – its state determining the readiness to the penetration through the pore. We then may build the construction analogous to the considered example, for which the bi-photonic control is more effective than the classical. 

We see that the application of the bi-photonic control in the different situations can raise the probability characteristic of efficiency comparatively with the classical control. Since the evolution of cell is the sequence of many steps, the probability characteristics of effectiveness will often multiply on all steps and the resulting effectiveness of the bi-photonic control can substantially exceed the classical analogue. 

The using of bi-photons for the establishing of the connections between the remote processors seems to be the most promising way of the development of the supercomputer technologies. This hybrid of classical silicon computer with the bi-photonic source is the most realistic embodiment of the idea of quantum computer nowadays. The development of this way can include also the more complex many photon states, for example, the entangled states of 4 and 6 photons that can help to represent more realistically the connection between processes simulated on the different computers. It is also possible to use the entangled states of particles with nonzero mass, for example, superconducting Cooper pairs, excited ion states in traps, or electrons in quantum dots. The software created on the basement of the collective behavior method makes possible to include easily entangled states of any type to our model. 

From the practical viewpoint, the using of entangled photon states is the best way to apply the miraculous quantum phenomenon of non-locality. I suppose that in the near perspective just this treatment of quantum computer is the most promising. The main area of applications for these computers will be the simulation of biochemistry. The further development of the idea of physical constructivism and its extension to the other types of complex systems is tightly connected with the development of the appropriate programming tools that requires the special efforts.

\begin{figure}
\centering
\caption{Constructive scheme of quantum computer}
\vspace{150mm}
\makebox[380mm][l]{\includegraphics{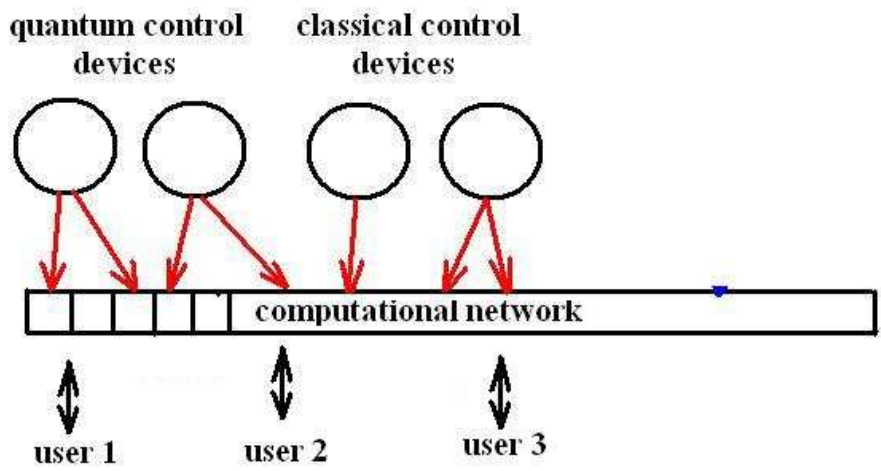}}%
\end{figure}

\section{Conclusion}

As the conclusion of the course, I summarize the main features of physical constructivism.

\begin{itemize}
\item Orientation to the dynamical scenarios instead of exact evaluations.
\item Pluralism: the admissibility of different scenarios, their comparison and selection. The direct including of the user into the model.
\item Possibility to consider mechanisms explicitly, on the level of samples. Scalability of the model. Preserving of the individuality of samples. Explicit form of the probability space. Explicit form of the entanglement as bonds between samples. 
\end{itemize}

The important component of the physical constructivism is the representation of decoherence as the fundamental factor, treated as the limitation of computational resources in the simulating computer. It brings the high price to the quantum entanglement effects obtained in experiments. Hence, the constructive approach to quantum computer differs from the Copenhagen (traditional) viewpoint. The quantum part of this computer is one of the controlling elements, e.g., is the kind of oracle. The constructive representation of quantum computer we schematically show at the picture.

In the standard quantum theory, the heuristic is the intermediate step, which serves to the obtaining of equations, which solution is the main aim. In the constructive quantum theory, equations are the auxiliary tool for the obtaining of the heuristic, which then serves as the basis for the computer simulation. In the simulation of standard situations reducible to one particle, the constructive approach gives only the braking of modeling.  It is effective only for the complex systems. What is the effect of constructivism for complex systems? It allows the direct manipulation with the entanglement as the physical phenomenon representing it as bonds. Standard Hilbert formalism has nothing similar. The presence of formal object – bond opens new possibilities for us. With this formal apparatus, we hope to obtain the description of many atom chemistry, e.g., to advance on the way of investigation of complex systems. 

Somebody could object to me, that it would not be the physics at all. I would like to avoid the discussion about terms. The essence is only valuable: if we intend to investigate complex systems, we must pass in physics to the constructivism. Only in this case we could expect the real advance. The division of sciences follows from the presence of peculiar phenomena in each of them, and by advance, we mean the expansion of the exact methods to these phenomena, e.g., the expansion of physics. One should clearly understand that this expansion cannot be free of charge, and the cost of constructivism is the lowest one.

\end{document}